\documentclass[12pt,a4paper]{article}
\usepackage[truedimen,margin=30mm]{geometry}

\usepackage{mathrsfs}
\usepackage{amssymb}
\usepackage{amsmath}
\usepackage{ascmac}
\usepackage{amsthm}
\usepackage{booktabs}
\usepackage{multirow}
\usepackage{tabularx}
\usepackage{threeparttable}
\usepackage{array}
\usepackage{graphicx}
\usepackage{bibunits}
\usepackage{natbib}
\usepackage{setspace}
\usepackage{dsfont}
\usepackage{times}
\usepackage[title]{appendix}
\usepackage{bbm}

\usepackage{hyperref}
\usepackage{comment}
\usepackage{mathtools}
\mathtoolsset{showonlyrefs}
\usepackage{tikz}
\usepackage{newtxtext,newtxmath}
\usetikzlibrary{arrows.meta,positioning,calc}
\usepackage{subcaption}

\usepackage{titlesec}
\titleformat*{\section}{\large\bfseries}
\titleformat*{\subsection}{\it}
\titleformat*{\subsubsection}{\it}

\newtheorem{df}{Definition}
\newtheorem{thm}{Theorem}

\newtheorem{prp}{Proposition}
\newtheorem{exm}{Example}

\newtheorem{rem}{Remark}
\DeclareMathOperator*{\argmax}{arg\,max}

\title{
The Covariate-Assisted Bayesian Intransitive \\
Bradley--Terry Model via Combinatorial \\
Hodge Theory}
\author{Hisaya Okahara$^{1\ast}$, Tomoyuki Nakagawa$^{2,4}$ and Shonosuke Sugasawa$^{3,4}$}
\date{}

\begin{document}

\maketitle
\doublespacing

\vspace{-1.3cm}
\begin{center}

\medskip
\noindent
$^1$ Department of Information Sciences, Tokyo University of Science \\
$^2$ School of Data Science, Meisei University \\
$^3$ Faculty of Economics, Keio University \\
$^4$ Center for Brain Science, RIKEN \\
$^\ast$ Corresponding author (Email : hisaya.okahara@gmail.com)
\end{center}

\vspace{1cm}
\begin{center}
{\bf \large Abstract}
\end{center}
Pairwise comparison data are widely used to recover latent rankings, yet the models in dominant use assume stochastic transitivity.
When preferences are in fact intransitive, a single scalar strength conflates genuine hierarchy with cycle-induced structure, biasing both the recovered ranking and any covariate effects attributed to it.
To address this limitation, we propose the Covariate-Assisted Bayesian Intransitive Bradley--Terry (CA-BIBT) model, which uses a combinatorial Hodge decomposition to resolve the latent match-up into identifiable and mutually orthogonal flows, attributing the component lying in the covariate-induced subspace to observed covariates and assigning the remaining components to the residuals.
A global-local shrinkage prior on the residual cycle-induced flow adapts the model from transitive to intransitive regimes without prespecifying the regime, and a Gibbs sampler yields, as posterior byproducts, calibrated uncertainty for each flow, the posterior probability of the level at which the entities are rankable, and two complementary decision summaries that remain well-defined under intransitivity.
In simulations, the CA-BIBT model recovers all flow components accurately with near-nominal coverage, and in applications to two animal dominance datasets, it distinguishes covariate-induced from residual cyclic dominance while quantifying posterior uncertainty, and demonstrates the practical utility of two complementary decision summaries.


\begin{bibunit}[agsm]
\section{Introduction}
\label{sec:introduction}

\medskip\noindent
Recovering a coherent ranking from pairwise comparisons is a central statistical problem, yet the probabilistic models most widely used for this task rest on an assumption of stochastic transitivity ($a$ likely beats $b$ and $b$ likely beats $c$ implies $a$ likely beats $c$).
The Bradley--Terry (BT) model \citep{bradley1952Rank} is the canonical example: it assigns each entity $i$ a latent score $s_i\in\mathbb{R}$ and models the probability that $i$ defeats $j$ as $p_{ij}=\sigma(s_i-s_j)$, where $\sigma(x)=(1+e^{-x})^{-1}$ is the logistic function, a parsimonious and interpretable specification that has become a cornerstone of the paired-comparison literature \citep{turner2012BradleyTerry, cattelan2012Models, hamilton2025Many}.

Transitivity is the assumption that does the work in those representations, and real comparison data frequently violate it.
Intransitive comparison structures are well documented across domains, from dominance hierarchies in animal behavior \citep{shoemaker1939Social,shizuka2012Social}, competitive games \citep{chen2016Modeling, smead2019Sports}, to inconsistent human judgments \citep{tversky1969Intransitivity, butler2018Predictably}.
When such structure is present, fitting a transitive model is misspecified, distorting both the recovered ranking and any inference drawn from it.

This misspecification is not benign and it is most severe for binary win--loss outcomes.
Under misspecification, maximum likelihood converges to the parameter value closest to the truth in the Kullback--Leibler (KL) sense, so the estimates are systematically pulled toward the best transitive approximation \citep{white1982Maximum, pukdee2026What}.
For cardinal data, complete and balanced experimental designs make the misspecified fit interpretable as recovering the score-difference component of the latent match-up, rather than as removing the misspecification itself \citep{singh2025Analysis}.
For binary data, however, the misspecified BT target is a KL projection taken after the nonlinear link, so experimental design no longer guarantees clean recovery of this component.

Existing tools characterize intransitivity only weakly, because they either measure it descriptively or model it without separating transitive strength from cycle-induced structure.
Descriptive counts of cyclic triads are not unique and need not be monotone in the distance from a transitive preference \citep{regenwetter2011Transitivity, mcdonald2013Comparative, sanjaya2022Measuring}.
Multidimensional embeddings encode cyclicity only implicitly, at the cost of interpretability and explicit quantification \citep{chen2016Modeling, duan2017Generalized, gu2021Intransitivity}, and recent variants target prediction rather than the quantification of intransitivity \citep{lee2025Pairwise, zhang2025BradleyTerry}.
Defining intransitivity as a displacement from a fitted BT model \citep{spearing2023Modeling} measures it against a baseline that is itself distorted by the bias above.

A common explanation nonetheless emerges in the literature: intransitivity reflects heterogeneity in the comparison itself, viewed either as a mixture of transitive orders \citep{gehrlein1983Condorcets, makhijani2019Parametric} or as an effect of matchup attributes, since similarity suppresses cyclicity \citep{cebra2023Similarity} and multiple skills or strategies may make transitivity unrealistic \citep{lee2025Pairwise}.
This suggests that covariates may account for part of the pairwise interaction, yet no existing framework uses them to characterize covariate-induced cyclic variation, to the best of our knowledge.
Covariate-assisted models such as \cite{springall1973Response} and \cite{fan2024Uncertainty} cannot serve this purpose, because their entity-specific covariates explain the same space as that of the BT model \citep{dong2025Statistical}.

We close this gap with the Covariate-Assisted Bayesian Intransitive Bradley--Terry (CA-BIBT) model, a logistic model for binary comparisons built on a combinatorial Hodge decomposition of the latent match-up \citep{jiang2011Statistical, lim2020Hodge}.
Following the BT convention of defining preferences for every pair, we work on the full comparison support, where the match-up splits uniquely into a gradient flow, representing a globally consistent transitive hierarchy, and a curl flow, capturing locally cyclic structure.
Into this geometry, the model places three mutually orthogonal flows, namely a covariate flow driven by pair-specific attributes, a residual gradient flow, and a residual curl flow, so that covariates explain whatever transitive and cycle-induced structure they can while the residuals absorb the rest.
We prove that the generating parameters are identifiable under explicit linear constraints (Theorem \ref{thm:identifiability_CA-BIBT}) and show that the model subsumes existing specifications \citep{bradley1952Rank, springall1973Response, fan2024Uncertainty}.

The posterior draws of the latent match-up support three inferential capabilities.
First, the framework provides calibrated uncertainty quantification for each flow component: a horseshoe shrinkage prior \citep{carvalho2010Horseshoe} on the residual curl flow regularizes potentially high-dimensional cycle-induced structure adaptively, letting one model span transitive to intransitive specifications.
Second, posterior probabilities for the stochastic transitivity (ST) classes \citep{fishburn1973Binary} arise as a byproduct, replacing separate hypothesis tests \citep{regenwetter2010Testing, shev2014Systemic} with direct probabilistic evidence of the level at which the entities are rankable.
Third, we introduce two complementary decision summaries that remain well-defined even under intransitivity: a blockwise ranking that reports how finely the entities can be ordered and a dominance graph that retains credible pairwise directions.
Both summaries are calibrated using a Bayesian false discovery rate (BFDR) to control the posterior expected proportion of erroneous directional claims at a prespecified tolerable level \citep{muller2006FDR}.
Posterior inference is carried out by a conditionally conjugate Gibbs sampler based on P\'{o}lya--Gamma data augmentation \citep{polson2013Bayesian}.

The rest of the article is organized as follows.
In Section \ref{sec:setup}, we introduce the problem setup and summarize the combinatorial Hodge theory underlying the proposed framework.
In Section \ref{sec:proposal}, we present the proposed CA-BIBT model, discuss identifiability, specify the Bayesian prior specification, describe the posterior computation algorithm, and show how several existing paired-comparison models arise as special cases.
In Section \ref{sec:quantification}, we develop posterior summaries for evaluating evidence for the ST classes and introduce complementary decision summaries for ranking and pairwise dominance assessment.
In Section \ref{sec:simulation}, we report simulation studies comparing the proposed method with existing approaches, and in Section \ref{sec:applications}, we apply it to two animal dominance datasets.
Finally, Section \ref{sec:discussion} concludes with a discussion and future directions.
Additional technical details, proofs, and supplementary simulation results are provided in the Supplementary Material, and \texttt{R} code implementing the proposed method is available at the GitHub repository \url{https://github.com/h-okahara/CA-BIBT}.

\section{Problem Setup}
\label{sec:setup}
Section~\ref{subsec:pairwise_comparisons} introduces the pairwise comparison setting, the match-up representation of preferences, and the stochastic transitivity (ST) classes that formalize a coherent ranking.
Section~\ref{subsec:HodgeDecomposition} reviews the combinatorial Hodge decomposition, which provides the geometric foundation on which the proposed model separates transitive from cycle-induced structure.

\subsection{Pairwise Comparisons}
\label{subsec:pairwise_comparisons}
Pairwise comparison data are summarized by an undirected comparison graph $\mathcal{G}=(\mathcal{V},\mathcal{E})$ with vertex set $\mathcal{V}=\{1,\dots,N\}$ and edge set $\mathcal{E} = \{\{i,j\} \mid n_{ij}>0 \}$, where $n_{ij}$ denotes the number of comparisons observed between entities $i$ and $j$.
For each pair, let $y_{ij}$ denote the number of times $i$ beats $j$ in $n_{ij}$ comparisons, and we assume $y_{ij} \sim \mathrm{Bin}(n_{ij},p_{ij})$ independently across pairs with the $n_{ij}$ comparisons within a pair conditionally independent and identically distributed.
In this work, we focus on binary win--loss outcomes and exclude ties for simplicity.

A convenient way to represent pairwise preferences is through a \emph{match-up function} $M\colon \mathcal{V}^2\to\mathbb{R}$ satisfying the skew-symmetry condition $M_{ij}=-M_{ji}$ where $M_{ij}=M(i,j)$.
We express the winning probabilities as
\begin{equation}
\label{eq:matchup-general}
    p_{ij}=F(M_{ij}),\quad i,j\in\mathcal{V},
\end{equation}
where $F$ is a symmetric and strictly increasing cumulative distribution function satisfying $F(x)+F(-x)=1$.
Under \eqref{eq:matchup-general}, we have $p_{ij}>1/2$ if and only if $M_{ij}>0$, so the sign of $M_{ij}$ determines the preferred direction of the comparison.

A classical transitive specification assumes the existence of a latent score vector $\boldsymbol{s}=(s_1,\ldots,s_N)^\top \in\mathbb{R}^N$ such that
\begin{equation}
\label{def:LST}
    p_{ij} = F(s_i-s_j),\quad i,j\in\mathcal{V}.
\end{equation}
This is referred to as the linear stochastic transitivity (LST) model \citep{oliveira2018New}, which includes the Bradley--Terry and Thurstone models as special cases \citep{bradley1952Rank, thurstone1927Method}.
While \eqref{def:LST} provides a fully transitive representation through latent scores, our interest here is to characterize when the collection of pairwise probabilities is at least compatible with some coherent overall ordering.

To this end, we recall a nested class of the ST conditions \citep{fishburn1973Binary}.
For every distinct $i,j,k\in\mathcal{V}$ such that $p_{ij}\geq 1/2$ and $p_{jk}\geq 1/2$, a set of comparison probabilities $P=\{p_{ij}\}_{i,j\in\mathcal{V}}$ satisfies
\begin{itemize}
    \item \emph{strong stochastic transitivity} (SST) if $p_{ik}\geq \max\{p_{ij},p_{jk}\}$,
    \item \emph{moderate stochastic transitivity} (MST) if $p_{ik}\geq \min\{p_{ij},p_{jk}\}$,
    \item \emph{weak stochastic transitivity} (WST) if $p_{ik}\ge 1/2$.
\end{itemize}
These conditions are strictly ordered, $\mathrm{SST}\Rightarrow \mathrm{MST}\Rightarrow \mathrm{WST}$, and the LST model \eqref{def:LST} implies SST.
Among them, WST is the weakest condition that guarantees compatibility with an overall ordering.
Define the binary relation $\succsim$ on $\mathcal{V}$ by $i \succsim j \ \Leftrightarrow \ p_{ij} \geq 1/2$.
Since $p_{ij}+p_{ji}=1$ for $i,j\in\mathcal{V}$, the relation $\succsim$ is reflexive and complete; under WST it is moreover transitive, so $\succsim$ is a complete preorder \citep[see][]{fishburn1973Binary, regenwetter2010Testing}.
Writing $\succ$ and $\sim$ for its strict and indifference parts, $\succsim$ induces an ordered partition of $\mathcal{V}$ into indifference classes, corresponding to a ranking with possible ties.
Because our inferential target is precisely such a coherent ordering, we record it as a property of $P$.
\begin{df}[\textit{Stochastic Transitivity}]
\label{def:ST}
\textup{
A set of pairwise probabilities $P=\{p_{ij}\}_{i,j\in\mathcal{V}}$ is \emph{stochastically transitive} if the induced relation $\succsim$ is a complete preorder on $\mathcal{V}$.
Equivalently, $P$ is stochastically transitive if and only if WST holds.
}
\end{df}

When no such complete preorder exists, we say that $P$ exhibits \emph{stochastic intransitivity} (SIT).
Under SIT no single ordered partition is compatible with all pairwise preferences, which motivates the decision summaries developed in Section~\ref{subsec:decision_summaries}.
\begin{exm}[\textit{Triadic configurations under SIT}]
\label{exm:SIT}
\textup{
Representative SIT configurations involve distinct $i,j,k\in\mathcal{V}$ with either 
(i) $i\succ j$, $j\succ k$, and $k\succ i$, or (ii) $i\sim j$, $j\succ k$, and $k\succ i$.
}
\end{exm}

The combinatorial Hodge decomposition reviewed next makes the source of intransitivity geometrically transparent, separating the transitive score-difference structure from a complementary cycle-induced component.
Because this decomposition is defined directly on the match-up scale, it does not depend on the particular choice of the link $F$.
In the remainder of the paper, we adopt the logistic link $F=\sigma$, as in the BT model, and build the proposed Bayesian inferential framework on this specification.
The quantitative connection between the latent match-up values and the ST classes above is established in Section~\ref{sec:quantification}.

\subsection{Combinatorial Hodge Decomposition}
\label{subsec:HodgeDecomposition}
Let $\mathcal{G}=(\mathcal{V},\mathcal{E})$ be the comparison graph and let $\mathcal{T} = \bigl\{\{i,j,k\} \in \binom{\mathcal{V}}{3} \mid \{i,j\},\{j,k\},\{k,i\} \in \mathcal{E} \bigr\}$ be its set of triangles.
We work with the clique complex of $\mathcal{G}$ restricted to dimension at most two, which augments the graph by filling each triangle with a $2$-simplex.
Thus the simplices consist of vertices $\mathcal{V}$, edges $\mathcal{E}$, and triangles $\mathcal{T}$.
The basic objects are a vertex function $s\colon \mathcal{V} \to \mathbb{R}$, an alternating edge flow $X\colon \mathcal{V}^2 \to\mathbb{R}$ with $X(i,j)=-X(j,i)$, and an alternating triangular flow $\Phi\colon \mathcal{V}^3 \to\mathbb{R}$ satisfying $\Phi(i,j,k)=\Phi(j,k,i)=\Phi(k,i,j) =-\Phi(i,k,j)=-\Phi(k,j,i)=-\Phi(j,i,k)$ for all $\{i,j,k\}\in\mathcal{T}$.

We denote the corresponding function spaces by $L^2(\mathcal{V})$, $L_\wedge^2(\mathcal{E})$, and $L_\wedge^2(\mathcal{T})$, respectively, and equip them with the standard $L^2$ inner products
\begin{gather}
    \langle s, t\rangle_\mathcal{V} = \sum_{i\in\mathcal{V}} s(i)t(i), \quad
    \langle X, Y\rangle_\mathcal{E} = \sum_{\{i,j\}\in\mathcal{E}} X(i,j)Y(i,j), \\
    \langle \Phi, \Psi\rangle_\mathcal{T} = \sum_{\{i,j,k\}\in\mathcal{T}} \Phi(i,j,k) \Psi(i,j,k).
\end{gather}

The combinatorial gradient operator $\mathrm{grad}\colon L^2(\mathcal{V}) \to L_\wedge^2(\mathcal{E})$ and the combinatorial curl operator $\mathrm{curl}\colon L_\wedge^2(\mathcal{E}) \to L_\wedge^2(\mathcal{T})$ are defined by
\begin{equation}
\label{def:operators}
\begin{aligned}
    (\mathrm{grad}\, s)(i,j) &= s(i) - s(j),\quad s \in L^2(\mathcal{V}),\\
    (\mathrm{curl}\, X)(i,j,k) &= X(i,j)+X(j,k)+X(k,i),\quad X \in L_\wedge^2(\mathcal{E}).
\end{aligned}
\end{equation}
Their adjoint operators $\mathrm{grad}^\ast\colon L_\wedge^2(\mathcal{E}) \to L^2(\mathcal{V})$ and $\mathrm{curl}^\ast\colon L_\wedge^2(\mathcal{T}) \to L_\wedge^2(\mathcal{E})$, characterized by the relations $\langle \mathrm{grad}\, s, X \rangle_\mathcal{E} = \langle s, \mathrm{grad}^\ast X \rangle_\mathcal{V}$ and $\langle \mathrm{curl}\, X, \Phi \rangle_\mathcal{T} = \langle X, \mathrm{curl}^\ast \Phi \rangle_\mathcal{E}$, are given explicitly by
\begin{equation}
\label{def:adjoints}
\begin{aligned}
    (\mathrm{grad}^\ast X)(i) &= \sum_{j:\{i,j\}\in\mathcal{E}} X(i,j),\quad X \in L_\wedge^2(\mathcal{E}) \\
    (\mathrm{curl}^\ast \Phi)(i,j) &= \sum_{k:\{i,j,k\}\in\mathcal{T}} \Phi(i,j,k),\quad \Phi \in L_\wedge^2(\mathcal{T}).
\end{aligned}
\end{equation}
Because $\mathrm{curl} \circ \mathrm{grad} = 0$, we have $\langle \mathrm{grad}\, s, \mathrm{curl}^\ast \Phi \rangle_\mathcal{E} = 0$ for all $s \in L^2(\mathcal{V})$ and $\Phi \in L_\wedge^2(\mathcal{T})$; that is, the gradient and curl flows are orthogonal in $L_\wedge^2(\mathcal{E})$.

In general, the \emph{combinatorial Hodge decomposition} takes the form
\begin{equation} \label{eq:Hodge}
    L_\wedge^2(\mathcal{E}) = 
    \lefteqn{\underbrace{\phantom{\mathrm{im}(\mathrm{grad})\ \oplus\ \ker(\Delta_1)}}_{\ker(\mathrm{curl})}}
    \mathrm{im}(\mathrm{grad})\ \oplus\
    \overbrace{\ker(\Delta_1)\ \oplus\ \mathrm{im}(\mathrm{curl}^\ast)}^{\ker(\mathrm{grad}^\ast)}
\end{equation}
where $\Delta_1 = \mathrm{grad}\circ\mathrm{grad}^\ast + \mathrm{curl}^\ast\circ\mathrm{curl}$.
The harmonic flow space $\ker(\Delta_1)$ consists of edge flows that lie in both $\ker(\mathrm{curl})$ and $\ker(\mathrm{grad}^\ast)$, whose dimension is called the first Betti number $\beta_1 = \dim \ker(\Delta_1)$.
More generally, $\beta_1=0$ if and only if every cycle in $\mathcal{G}$ lies in the span of triangle boundaries; chordality is a sufficient condition, and the complete graph is a special case \citep{lim2020Hodge}.

In pairwise comparison models such as the BT model, latent preferences are modeled for every pair of entities, regardless of whether that pair is observed in the data.
Accordingly, although the observed comparison graph may be incomplete, the latent comparison structure is naturally represented by the complete graph.
We adopt the same perspective in the proposed model, so the decomposition reduces to
\begin{equation}
\label{eq:Hodge_complete}
  L_\wedge^2(\mathcal{E}) = \mathrm{im}(\mathrm{grad}) \oplus \mathrm{im}(\mathrm{curl}^\ast).
\end{equation}
Consequently, every edge flow decomposes uniquely into a \emph{gradient flow}, representing a globally consistent hierarchical structure, and a \emph{curl flow}, capturing cycle-induced patterns.
This additive decomposition acts on the match-up scale, and it is on this scale that the transitive and cyclic components are orthogonal. 
Because the latent comparison support is the complete graph, the harmonic space is trivial ($\beta_1 = 0$), so no harmonic component arises in the latent match-up; the case where the observed support is incomplete is discussed in Remark~\ref{rem:obs_graph} and Supplementary Material~S4.2.

Fixing the canonical orientations induced by the vertex labels and ordering the oriented simplices lexicographically identifies $L^2(\mathcal V)\simeq \mathbb R^N$, $L_\wedge^2(\mathcal E)\simeq \mathbb R^{|\mathcal E|}$ and $L_\wedge^2(\mathcal T)\simeq \mathbb R^{|\mathcal T|}$, where for the complete graph $|\mathcal{E}|=\binom{N}{2}$ and $|\mathcal{T}|=\binom{N}{3}$.
Scalar potentials, edge flows, and triangular flows are then coordinate vectors $\boldsymbol{s}\in\mathbb{R}^N$, $\boldsymbol{M}\in\mathbb{R}^{|\mathcal E|}$, and $\boldsymbol{\Phi}\in\mathbb{R}^{|\mathcal T|}$, and $\mathrm{grad}$ and $\mathrm{curl}$ are represented by matrices $G\in\mathbb{R}^{|\mathcal E|\times N}$ and $C\in\mathbb{R}^{|\mathcal T|\times|\mathcal E|}$.
Under the standard $L^2$ inner products their adjoints are $G^\top$ and $C^\top$.

\section{The Covariate-Assisted Bayesian Intransitive Bradley--Terry Model}
\label{sec:proposal}
The remainder of this section is organized as follows.
Section~\ref{subsec:model} introduces the CA-BIBT model and establishes its identifiability under Hodge-theoretic linear constraints.
Section~\ref{subsec:prior} specifies prior distributions for all parameters.
Section~\ref{subsec:Gibbs} presents the posterior computation algorithm via Gibbs sampling.
Section~\ref{subsec:related_work} positions the CA-BIBT model relative to existing paired-comparison models.

\subsection{The Proposed Model}
\label{subsec:model}
Since pairwise preferences are represented by an alternating latent match-up vector $\boldsymbol{M}\in\mathbb{R}^{|\mathcal{E}|}$, the latent comparison structure can be viewed as an edge flow.
In this paper, the combinatorial Hodge decomposition is applied to the latent comparison support, which we take to be the complete graph on $\mathcal{V}$.
Under this complete latent support, the latent structure of $\boldsymbol{M}$ can be decomposed into a gradient component in $\mathrm{im}(\mathrm{grad})$ and a curl component in $\mathrm{im}(\mathrm{curl}^\ast)$.
To incorporate observed covariate information, we introduce additional notation.

For each oriented pair $(i,j)$, let $\boldsymbol{x}_{ij} \in \mathbb{R}^d$ denote an edge-specific covariate vector satisfying the antisymmetry condition $\boldsymbol{x}_{ji} = -\boldsymbol{x}_{ij}$.
Let
\begin{equation}
    X_E = \bigl(\boldsymbol{x}_{12}, \ldots, \boldsymbol{x}_{(N-1)N}\bigr) \in\mathbb{R}^{d \times |\mathcal{E}|}
\end{equation}
be the covariate design matrix, and assume that $X_E$ has full row rank.
For a coefficient vector $\boldsymbol{\beta} \in \mathbb{R}^d$, the product $X_E^\top \boldsymbol{\beta} \in\mathbb{R}^{|\mathcal{E}|}$ defines an edge flow, which we refer to as the \emph{covariate flow}.

Under the logistic link, the winning probabilities are $p_{ij}=\sigma(M_{ij})$, and we define the \emph{Covariate-Assisted Bayesian Intransitive Bradley--Terry} (CA-BIBT) model by specifying the latent match-up vector as
\begin{equation}
\label{def:CA-BIBT}
    \boldsymbol{M} = G\boldsymbol{s} + C^\top\boldsymbol{\Phi} + X_E^\top\boldsymbol{\beta}.
\end{equation}
The observation model is thus $y_{ij}\sim\mathrm{Bin}(n_{ij},\sigma(M_{ij}))$, where $G\boldsymbol{s}$ is the gradient flow, $C^\top\boldsymbol{\Phi}$ is the curl flow, and $X_E^\top\boldsymbol{\beta}$ is the covariate flow.

By the combinatorial Hodge decomposition \eqref{eq:Hodge_complete}, every edge flow admits a unique decomposition into a gradient component $\boldsymbol{M}_{\mathrm{grad}}\in\mathrm{im}(\mathrm{grad})$ and a curl component $\boldsymbol{M}_{\mathrm{curl}}\in\mathrm{im}(\mathrm{curl}^\ast)$ \citep[see][]{bhatia2013HelmholtzHodge, strang2022Network}.
However, the parameters $(\boldsymbol{s}, \boldsymbol{\Phi}, \boldsymbol{\beta})$ generating these components are not uniquely determined, since $\ker(\mathrm{grad})$ and $\ker(\mathrm{curl}^\ast)$ are nontrivial.
Furthermore, the covariate flow $X_E^\top\boldsymbol{\beta}$ generally contains both gradient and curl components.
These sources of non-uniqueness render the CA-BIBT model unidentifiable without further restrictions.

To achieve identifiability, we impose two sets of constraints.
Let
\begin{equation}
    K = \dim\,\mathrm{im}(\mathrm{curl}^\ast) = |\mathcal{E}| - |\mathcal{V}| + 1 
    = \tbinom{N-1}{2},
\end{equation}
and define the kernel spaces as
\begin{equation}
    \ker(\mathrm{grad}) = \mathrm{span}(\boldsymbol{1}), \qquad
    \ker(\mathrm{curl}^\ast) = \bigl\{ A\boldsymbol{w} \mid 
    \boldsymbol{w} \in \mathbb{R}^{|\mathcal{T}|-K} \bigr\},
\end{equation}
where $A \in \mathbb{R}^{|\mathcal{T}| \times (|\mathcal{T}| - K)}$ is a 
full-column-rank matrix whose columns form a basis for $\ker(\mathrm{curl}^\ast)$.

The first set addresses the non-uniqueness of $\boldsymbol{s}$ and $\boldsymbol{\Phi}$ within their respective subspaces.
Let $\boldsymbol{v} \in \mathbb{R}^N$ satisfy $\boldsymbol{v}^\top\boldsymbol{1} \neq 0$ and $V \in \mathbb{R}^{|\mathcal{T}| \times (|\mathcal{T}| - K)}$ satisfy $\det(V^\top A) \neq 0$.
We impose the linear constraints $\boldsymbol{v}^\top \boldsymbol{s} = 0$ and $V^\top \boldsymbol{\Phi} = \boldsymbol{0}$, which remove the indeterminacies arising from $\ker(\mathrm{grad})$ and $\ker(\mathrm{curl}^\ast)$, respectively.
Standard choices include the sum-to-zero constraint $\sum_{i=1}^N s_i = 0$ and the reference constraint $s_1 = 0$ for the score parameter $\boldsymbol{s}$.

The second set ensures that the gradient and curl flows are not confounded with the covariate flow.
Specifically, we impose the orthogonality conditions $X_E G\boldsymbol{s} = \boldsymbol{0}$ and $X_E C^\top \boldsymbol{\Phi} = \boldsymbol{0}$, which guarantee that $G\boldsymbol{s}$ and $C^\top\boldsymbol{\Phi}$ are orthogonal to the column space of $X_E^\top$, and hence carry no component lying in the covariate flow space.

The following theorem establishes that these constraints together ensure identifiability of the CA-BIBT model.
\begin{thm}[\textit{Identifiability of the CA-BIBT model}]
\label{thm:identifiability_CA-BIBT}
\textup{
Let 
\begin{equation}
    \boldsymbol{M} = G\boldsymbol{s} + C^\top\boldsymbol{\Phi} + X_E^\top\boldsymbol{\beta},
\end{equation}
with $\boldsymbol{s} \in \mathbb{R}^N$, $\boldsymbol{\Phi} \in \mathbb{R}^{|\mathcal{T}|}$, and $\boldsymbol{\beta} \in \mathbb{R}^d$.
Suppose that $\boldsymbol{v}^\top\boldsymbol{1}\neq 0$, $\det(V^\top A)\neq 0$, and $X_E$ has full row rank.
Define the constrained parameter space
\begin{equation}
    \Theta (\boldsymbol{v}, V, X_E) = \left\{ (\boldsymbol{s},\boldsymbol{\Phi},\boldsymbol{\beta}) \ \middle| \  
    \boldsymbol{v}^\top\boldsymbol{s}=0,\ V^\top\boldsymbol{\Phi}=\boldsymbol{0},\ 
    X_E G\boldsymbol{s} = \boldsymbol{0},\ X_E C^\top\boldsymbol{\Phi} = \boldsymbol{0} \right\}.
\end{equation}
Then the CA-BIBT model is identifiable over $\Theta (\boldsymbol{v},V,X_E)$; equivalently, the mapping $(\boldsymbol{s},\boldsymbol{\Phi},\boldsymbol{\beta}) \in\Theta(\boldsymbol{v},V,X_E) \mapsto \{\sigma(M_{ij})\}_{\{i,j\}\in\mathcal{E}}$ is injective.
}
\end{thm}

\begin{rem}[\textit{Flow Contribution Ratios}]
\label{rem:flow-contribution}
\textup{
The flow decomposition also provides posterior attribution summaries for the latent match-up vector.
Let $\boldsymbol{M}_{g,r}=G\boldsymbol{s}$, $\boldsymbol{M}_{c,r}=C^\top\boldsymbol{\Phi}$, and $\boldsymbol{M}_{x}=X_E^\top\boldsymbol{\beta}$.
Let $P_\mathrm{grad}=G(G^\top G)^+G^\top$ and $P_\mathrm{curl}=C^\top(CC^\top)^+C$, where $+$ denotes the Moore--Penrose inverse, denote the orthogonal projections onto the gradient and curl flow spaces, respectively.
We then define the gradient and curl components of the covariate flow by $\boldsymbol{M}_{g,x}=P_\mathrm{grad}\boldsymbol{M}_{x}$ and $\boldsymbol{M}_{c,x}=P_\mathrm{curl}\boldsymbol{M}_{x}$.
Thus, under the identifiability constraints in Theorem~\ref{thm:identifiability_CA-BIBT}, the four components $\boldsymbol{M}_{g,r}$, $\boldsymbol{M}_{c,r}$, $\boldsymbol{M}_{g,x}$, and $\boldsymbol{M}_{c,x}$ form a mutually orthogonal decomposition of $\boldsymbol{M}$.
}

\textup{
For $\|\boldsymbol{M}\|_2 >0$, we define the component-wise contribution ratios as
\begin{equation}
    R_{g,r} = \frac{\|\boldsymbol{M}_{g,r} \|_2^2}{\|\boldsymbol{M}\|_2^2},\quad
    R_{c,r} = \frac{\|\boldsymbol{M}_{c,r} \|_2^2}{\|\boldsymbol{M}\|_2^2},\quad
    R_{g,x} = \frac{\|\boldsymbol{M}_{g,x} \|_2^2}{\|\boldsymbol{M}\|_2^2},\quad
    R_{c,x} = \frac{\|\boldsymbol{M}_{c,x} \|_2^2}{\|\boldsymbol{M}\|_2^2}.
\end{equation}
The total gradient and curl contribution ratios, and the covariate contribution ratio, are given by $R_g = R_{g,r} + R_{g,x}$, $R_c = R_{c,r} + R_{c,x}$, and $R_x = R_{g,x} + R_{c,x}$, respectively.
When the covariate contribution within each Hodge subspace is of interest, we also use $R_{x \mid g} = R_{g,x}/R_g$ and $R_{x \mid c} = R_{c,x}/R_c$.
These quantities are used only as posterior attribution summaries; they do not define the ST class, which is determined by the local conditions in Section~\ref{subsec:measures}.
}
\end{rem}

\begin{rem}[\textit{Recoverability under Incomplete Observation}]
\label{rem:obs_graph}
\textup{
When the observed comparison graph is incomplete, the likelihood is supported only on the observed edges.
Hence, latent match-up values on unobserved edges are not identifiable from the observed data alone, and posterior inference for such edges should be interpreted as model-based extrapolation.
In particular, direct recovery of the full latent match-up structure requires an observation design in which all pairs are eventually observed.
To assess the empirical robustness of the CA-BIBT model to graph incompleteness, Supplementary Material~S4.2 examines how posterior accuracy changes as the missing rate varies.
The results show that covariate information helps mitigate the effect of missing edges, although it does not remove the fundamental non-identifiability of unobserved edge-level structure.
}
\end{rem}

\subsection{Bayesian Prior Specification}
\label{subsec:prior}
For posterior computation, we specialize the identifiability constraints in Theorem~\ref{thm:identifiability_CA-BIBT} to the choices $\boldsymbol{v} = \boldsymbol{1}$ and $V = A$.
Under these choices, the constrained parameter space reduces to
\begin{equation}
\label{sup:parameter_space}
    \boldsymbol{s} \in \ker
    \begin{pmatrix}
    \boldsymbol{1}^\top \\ 
    X_E G 
    \end{pmatrix},\quad
    \boldsymbol{\Phi} \in \ker
    \begin{pmatrix} A^\top \\
    X_E C^\top 
    \end{pmatrix},
\end{equation}
of dimensions $q_u = N-1-d_s$ and $q_z = K-d_c$, respectively, where $d_s = \mathrm{rank}(X_E G)$ and $d_c = \mathrm{rank}(X_E C^\top)$.

To handle these constraints in the Gibbs sampler, we adopt a constraint-free reparameterization via orthonormal bases of \eqref{sup:parameter_space}.
Let $B_g \in \mathbb{R}^{N\times q_u}$ and $B_c \in \mathbb{R}^{|\mathcal{T}|\times q_z}$ denote orthonormal bases.
We then reparameterize
\begin{equation}
\label{sup:reparameterization}
    \boldsymbol{s} = B_g \boldsymbol{u},\quad
    \boldsymbol{\Phi} = B_c \boldsymbol{z},
\end{equation}
with $\boldsymbol{u}\in\mathbb{R}^{q_u}$ and $\boldsymbol{z}\in\mathbb{R}^{q_z}$.
Under \eqref{sup:reparameterization}, the identifiability constraints are satisfied, and the match-up vector admits the reduced form
\begin{equation}
\label{eq:CA-BIBT_reparam}
    \boldsymbol{M} = D_g \boldsymbol{u} + D_c \boldsymbol{z} + X_E^\top \boldsymbol{\beta},
\end{equation}
where $D_g = G B_g \in \mathbb{R}^{|\mathcal{E}|\times q_u}$ and $D_c = C^\top B_c \in \mathbb{R}^{|\mathcal{E}|\times q_z}$ are the corresponding design matrices.

We now specify prior distributions on the unconstrained parameters $(\boldsymbol{u}, \boldsymbol{z}, \boldsymbol{\beta})$ in \eqref{eq:CA-BIBT_reparam}.
For the covariate coefficient vector $\boldsymbol{\beta} = (\beta_1,\ldots,\beta_d)^\top$, we assign
\begin{equation}
    \beta_\ell \mid \sigma_\beta^2 \sim N(0, \sigma_\beta^2),\quad
    \ell = 1,\ldots,d,
\end{equation}
with $\sigma_\beta^2 \sim \mathrm{IG}(1/2, 1/2)$.
This Gaussian prior on $\boldsymbol{\beta}$ treats the covariate flow $X_E^\top\boldsymbol{\beta}$ as an integrated contribution, whose gradient and curl components are naturally recovered from posterior samples.

For the residual score coefficient vector $\boldsymbol{u} = (u_1,\ldots,u_{q_u})^\top$, we assume
\begin{equation}
    u_i \mid \sigma_u^2 \sim N(0, \sigma_u^2),\quad
    i=1,\ldots,q_u,
\end{equation}
with $\sigma_u^2 \sim \mathrm{IG}(1/2, 1/2)$.

For the residual curl coefficient vector $\boldsymbol{z} = (z_1,\ldots,z_{q_z})^\top$, which typically lies in a high-dimensional space and is expected to be low-dimensional, we adopt the Horseshoe prior \citep{carvalho2009Handling, carvalho2010Horseshoe}:
\begin{gather}
    \boldsymbol{z} \mid \tau, \{\lambda_\ell\} \sim N(\boldsymbol{0}, W), \\
    \lambda_\ell \sim C^+(0,1),\quad 
    \tau \sim C^+(0,1),\quad \ell = 1,\ldots,q_z,
\end{gather}
where $W = \mathrm{diag}(\tau^2\lambda_1^2, \ldots, \tau^2\lambda_{q_z}^2)$.
The global parameter $\tau$ controls the overall degree of shrinkage, whereas the local parameters $\{\lambda_\ell\}$ allow individual residual curl directions to escape shrinkage when strongly supported by the data.
Here, all prior components are taken to be mutually independent.

\subsection{Posterior Computation Algorithm}
\label{subsec:Gibbs}
To facilitate Gibbs sampling under the reduced reparameterization introduced in \eqref{eq:CA-BIBT_reparam}, we employ the hierarchical representation of the Horseshoe prior \citep{makalic2016Simple}.
The local and global scale parameters $\lambda_\ell \sim C^+(0,1)$ and $\tau \sim C^+(0,1)$ can be expressed hierarchically as
\begin{equation}
    \lambda_\ell^2 \mid \nu_\ell \sim \mathrm{IG} \left(\tfrac{1}{2}, \tfrac{1}{\nu_\ell}\right),\quad
    \tau^2 \mid \xi \sim \mathrm{IG} \left(\tfrac{1}{2}, \tfrac{1}{\xi}\right),\quad
    \nu_\ell, \xi \sim \mathrm{IG} \left(\tfrac{1}{2},1\right),\quad
    \ell=1,\ldots,q_z.
\end{equation}
By introducing the auxiliary variables $\{\nu_\ell\}$ and $\xi$, both the local and the global scales can be sampled from inverse-gamma distributions, thereby enabling straightforward Gibbs updates.

The joint posterior distribution of $(\boldsymbol{\beta}, \sigma_\beta^2, \boldsymbol{u}, \sigma_u^2, \boldsymbol{z}, \{\lambda_\ell^2\}, \tau^2, \{\nu_\ell\}, \xi)$ is given by
\begin{equation} \label{eq:pos_CA-BIBT}
\begin{split}
& p \left(\boldsymbol{\beta}, \sigma_\beta^2, \boldsymbol{u}, \sigma_u^2, \boldsymbol{z}, \{\lambda_\ell^2\}, \tau^2, \{\nu_\ell\}, \xi \ \middle|\ Y\right) \\
&\quad \propto 
\prod_{i<j} \frac{\{\exp(M_{ij})\}^{y_{ij}}}{\{1+\exp(M_{ij})\}^{n_{ij}}}
\phi \left(\boldsymbol{\beta}; \boldsymbol{0}, \sigma_\beta^2 I\right)
g \left(\sigma_\beta^2; \tfrac{1}{2}, \tfrac{1}{2}\right) 
\phi\!\left(\boldsymbol{u}; \boldsymbol{0}, \sigma_u^2 I\right)
g \left(\sigma_u^2; \tfrac{1}{2}, \tfrac{1}{2}\right) \\
&\qquad \times
\phi\left(\boldsymbol{z}; \boldsymbol{0}, W\right) 
\prod_{\ell=1}^{q_z} g \left(\lambda_\ell^2; \tfrac{1}{2}, \tfrac{1}{\nu_\ell}\right)
g\left(\nu_\ell; \tfrac{1}{2}, 1\right)
g\left(\tau^2; \tfrac{1}{2}, \tfrac{1}{\xi}\right)
g\left(\xi; \tfrac{1}{2}, 1\right),
\end{split}
\end{equation}
where $M_{ij}$ is defined through \eqref{def:CA-BIBT}, $\phi(\boldsymbol{x}; \boldsymbol{a},B)$ denotes the density of a normal distribution with mean $\boldsymbol{a}$ and covariance matrix $B$, and $g(x;a,b)$ denotes the density of an inverse-gamma distribution with shape $a$ and scale $b$.

To generate posterior samples from \eqref{eq:pos_CA-BIBT}, we employ P\'{o}lya--Gamma data augmentation \citep{polson2013Bayesian}.
For each likelihood term, we use the identity
\begin{equation}
    \frac{\{\exp(M_{ij})\}^{y_{ij}}}{\{1+\exp(M_{ij})\}^{n_{ij}}} 
    = 2^{-n_{ij}} \exp(\kappa_{ij}M_{ij}) \int_0^\infty \exp\left(-\frac{\omega_{ij}}{2}M_{ij}^2\right) f_\mathrm{PG}(\omega_{ij}) d\omega_{ij},
\end{equation}
where $\kappa_{ij} = y_{ij} - n_{ij}/2$ and $f_\mathrm{PG}(\omega_{ij})$ is the density of the P\'{o}lya--Gamma distribution $\mathrm{PG}(n_{ij}, 0)$.
Introducing latent variables $\omega_{ij} \sim \mathrm{PG}(n_{ij}, M_{ij})$ for each pair $(i,j)$ and collecting them into $\boldsymbol{\omega} = (\omega_{12}, \ldots, \omega_{(N-1)N})^\top$, $\Omega = \mathrm{diag}(\boldsymbol{\omega})$, and $\boldsymbol{\kappa} = (\kappa_{12}, \ldots, \kappa_{(N-1)N})^\top$, the full conditional distributions of $\boldsymbol{\omega}$, $\boldsymbol{\beta}$, $\sigma_\beta^2$, $\boldsymbol{u}$, $\sigma_u^2$, $\boldsymbol{z}$, $\{\lambda_\ell^2\}$, $\tau^2$, $\{\nu_\ell\}$, and $\xi$ take familiar conjugate forms.
The detailed sampling procedures are described below.

\begin{itemize}
\item[-] \textbf{(Update of $\boldsymbol{\omega}$)}\ 
For $1 \le i < j \le N$, generate $\omega_{ij}$ from $\mathrm{PG}(n_{ij}, M_{ij})$.

\item[-] \textbf{(Update of $\boldsymbol{\beta}$)}\ 
Generate $\boldsymbol{\beta}$ from $N(A_\beta B_\beta, A_\beta)$, where
\begin{equation*}
    A_\beta = \left(\sigma_\beta^{-2} I + X_E \Omega X_E^\top\right)^{-1},\quad
    B_\beta = X_E \left\{\boldsymbol{\kappa} - \Omega\!\left(D_g \boldsymbol{u} + D_c \boldsymbol{z}\right)\right\}.
\end{equation*}

\item[-] \textbf{(Update of $\sigma_\beta^2$)}\ 
Generate $\sigma_\beta^2$ from $\mathrm{IG}\left(a_\sigma, b_\sigma \right)$, where 
\begin{equation}
     a_\sigma = \frac{1+d}{2},\quad
     b_\sigma = \frac{1+\boldsymbol{\beta}^\top \boldsymbol{\beta}}{2}.
\end{equation}

\item[-] \textbf{(Update of $\boldsymbol{u}$)}\ 
Generate $\boldsymbol{u}$ from $N(A_u B_u, A_u)$, where
\begin{equation*}
    A_u = \left(\sigma_u^{-2} I + D_g^\top \Omega D_g\right)^{-1},\quad
    B_u = D_g^\top \left\{\boldsymbol{\kappa} - \Omega\!\left(D_c \boldsymbol{z} + X_E^\top \boldsymbol{\beta}\right)\right\}.
\end{equation*}

\item[-] \textbf{(Update of $\sigma_u^2$)}\ 
Generate $\sigma_u^2$ from $\mathrm{IG}\left(a_u, b_u\right)$, where
\begin{equation}
    a_u = \frac{1+q_u}{2},\quad
    b_u = \frac{1+\boldsymbol{u}^\top \boldsymbol{u}}{2}.
\end{equation}

\item[-] \textbf{(Update of $\boldsymbol{z}$)}\ 
Generate $\boldsymbol{z}$ from $N(A_z B_z, A_z)$, where
\begin{equation*}
    A_z = \left(W^{-1} + D_c^\top \Omega D_c\right)^{-1},\quad
    B_z = D_c^\top \left\{\boldsymbol{\kappa} - \Omega\!\left(D_g \boldsymbol{u} + X_E^\top \boldsymbol{\beta}\right)\right\}.
\end{equation*}

\item[-] \textbf{(Update of $\lambda_\ell^2$)}\ 
For $\ell = 1, \ldots, q_z$, generate $\lambda_\ell^2$ from $\mathrm{IG} (a_\lambda^{(\ell)}, b_\lambda^{(\ell)})$.
\begin{equation}
    a_\lambda^{(\ell)} = 1,\quad
    b_\lambda^{(\ell)} = \frac{1}{\nu_\ell} + \frac{z_\ell^2}{2\tau^2}
\end{equation}

\item[-] \textbf{(Update of $\tau^2$)}\ 
Generate $\tau^2$ from $\mathrm{IG}(a_\tau,b_\tau)$.
\begin{equation*}
    a_\tau = \frac{q_z + 1}{2},\quad
    b_\tau = \frac{1}{\xi} + \frac{1}{2} \sum_{\ell=1}^{q_z} \frac{z_\ell^2}{\lambda_\ell^2}.
\end{equation*}

\item[-] \textbf{(Update of $\nu_\ell$)}\ 
For $\ell = 1, \ldots, q_z$, generate $\nu_\ell$ from $\mathrm{IG}(a_\nu^{(\ell)}, b_\nu^{(\ell)})$, where $a_\nu^{(\ell)} = 1$ and $b_\nu^{(\ell)} = 1 + 1/\lambda_\ell^2$.

\item[-] \textbf{(Update of $\xi$)}\ 
Generate $\xi$ from $\mathrm{IG}(a_\xi, b_\xi)$, where $a_\xi = 1$ and $b_\xi = 1 + 1/\tau^2$.
\end{itemize}

The above Gibbs sampling algorithm can be easily implemented since most of the full conditional distributions are familiar due to the P\'{o}lya--Gamma data augmentation.

\subsection{Related Work}
\label{subsec:related_work}
The CA-BIBT model contains several representative paired-comparison models as special cases, thereby providing a unified view of existing formulations.
Table~\ref{tab:model-summary} summarizes which components are present in each model specification.
The BT model, the covariate-only structured BT model \citep{springall1973Response}, and the CARE model \citep{fan2024Uncertainty} generate curl-free edge flows, with the latter two using entity-specific covariates (see Supplementary Material~S2).
The proposed models instead admit curl components; the BIBT model is the covariate-free special case of the CA-BIBT model (see Supplementary Material~S1).
For a comprehensive review of covariate extensions in paired-comparison models, see \cite{cattelan2012Models, caron2012Efficient, li2022Bayesian}.

The CA-BIBT model is distinguished from recent covariate extensions by its treatment of static edge-specific covariates.
A separate line of work instead relaxes the orthogonality restriction between covariate and latent effects by adopting a dynamic score construction.
\cite{dong2025Statistical} extend the Plackett--Luce framework with comparison-specific scores of the form $s_i + X_{e,i}^\top \boldsymbol{\beta}$, where $X_{e,i} \in\mathbb{R}^d$ denotes the covariates of entity~$i$ in comparison hyperedge $e \subseteq \mathcal{V}$, thereby accommodating dynamic, time-varying covariates such as home advantage, rather than static edge-specific covariates.
Their goal, however, is not to quantify intransitivity.
To the best of our knowledge, the CA-BIBT model is the first paired-comparison model for binary outcomes that unifies covariate-induced and residual curl flows within the Hodge-theoretic framework under explicit identifiability constraints.

\begin{table}[htbp]
\centering
\caption{Summary of representative paired-comparison models within the proposed framework.
The match-up column gives the corresponding latent edge-flow specification, and the remaining columns indicate whether each gradient or curl component is allowed by the model.
Note that $X=(\boldsymbol{x}_1,\ldots,\boldsymbol{x}_N)\in\mathbb{R}^{d\times N}$ is the entity-specific covariate matrix.}
\label{tab:model-summary}
\begin{threeparttable}
\begin{tabular}{cccccc}
\toprule
& & \multicolumn{2}{c}{Gradient Flow Space} & \multicolumn{2}{c}{Curl Flow Space} \\
\cmidrule(lr){3-4} \cmidrule(lr){5-6}
Model & Match-up & Latent$^\ddagger$ & Covariate & Latent & Covariate \\
\midrule
BT  & $G\boldsymbol{s}$
    & \checkmark & & & \\
Structured BT$^\dagger$ & $G X^\top \boldsymbol{\beta}$
    & & \checkmark & & \\
CARE & $G(\boldsymbol{s} + X^\top \boldsymbol{\beta})$
    & \checkmark & \checkmark & & \\
\textbf{BIBT} & $G\boldsymbol{s}+C^\top\boldsymbol{\Phi}$
    & \checkmark & & \checkmark & \\
\textbf{CA-BIBT} & $G\boldsymbol{s}+C^\top\boldsymbol{\Phi}+X_E^\top\boldsymbol{\beta}$
    & \checkmark & \checkmark & \checkmark & \checkmark \\
\bottomrule
\end{tabular}
\begin{tablenotes}
\footnotesize
\item[$\dagger$] Used here to denote the covariate-only structured BT specification.
\item[$\ddagger$] ``Latent'' refers to a free latent component, whereas ``Covariate'' refers to a component induced by observed covariates.
\end{tablenotes}
\end{threeparttable}
\end{table}

\section{Quantification of Stochastic Transitivity and Decision Making}
\label{sec:quantification}
Section~\ref{subsec:measures} introduces posterior summaries of intransitivity and posterior evidence for the ST classes.
Section~\ref{subsec:decision_summaries} then considers how a comparison graph may be summarized for ranking-based decision making even under the SIT class.
Let $\Pi(\cdot \mid Y)$ denote the posterior probability measure induced by $p(\cdot \mid Y)$.

\subsection{Posterior Evidence for the Stochastic Transitivity Class}
\label{subsec:measures}
We now construct posterior summaries for the ST classes introduced in Section~\ref{subsec:pairwise_comparisons}.
Because $F$ is strictly increasing and symmetric, each ST condition on the winning probabilities transfers without change to the match-up values $\{M_{ij}\}$; the resulting characterization is therefore link-free, holding for any symmetric, strictly increasing $F$ rather than for the logistic link alone.
This invariance lets us recast the ST conditions, originally stated on $P=\{p_{ij}\}_{i,j\in\mathcal{V}}$, as local conditions on the match-up values.

The ST classes can be characterized locally on triangular substructures through the curl operator.
For each ordered triplet $(i,j,k)$ of distinct vertices, we obtain $\mathcal{C}_{ijk} = (\mathrm{curl}\, M)(i,j,k) = M_{ij} + M_{jk} + M_{ki}$.
This quantity provides a local diagnostic of cyclic structure on the triad $\{i,j,k\}$, and the sign determines the orientation of the induced cycle, while its magnitude reflects the strength of the local departure from a curl-free, transitive representation.
Because the gradient flow is curl-free, $\mathcal{C}_{ijk}$ only depends on its curl component.

We use this quantity to evaluate posterior evidence for the ST classes.
For any posterior draw of $\boldsymbol{M}$, consider ordered triplets satisfying $M_{ij}\geq 0$ and $M_{jk}\geq 0$.
The ST conditions then impose lower bounds on $M_{ik}=M_{ij}+M_{jk}-\mathcal{C}_{ijk}$; equivalently, they impose upper bounds on the corresponding $\mathcal{C}_{ijk}$.
Define
\begin{align}
    V_\mathrm{S} &= \max \left\{ \mathcal{C}_{ijk} - \min(M_{ij}, M_{jk}) \mid M_{ij}\geq0,\ M_{jk}\geq0 \right\}, \\
    V_\mathrm{M} &= \max \left\{ \mathcal{C}_{ijk} - \max(M_{ij}, M_{jk}) \mid M_{ij}\geq0,\ M_{jk}\geq0 \right\}, \\
    V_\mathrm{W} &= \max \left\{ \mathcal{C}_{ijk} - (M_{ij} + M_{jk}) \mid M_{ij}\geq0,\ M_{jk}\geq0 \right\}.
\end{align}
Then SST, MST, WST, and SIT hold if and only if $V_\mathrm{S}\leq 0$, $V_\mathrm{M}\leq 0$, $V_\mathrm{W}\leq 0$, and $V_\mathrm{W}> 0$, respectively.
We therefore define the posterior probabilities 
\begin{equation}
    \pi_\mathrm{S} = \Pi(V_\mathrm{S}\leq 0 \mid Y),\quad
    \pi_\mathrm{M} = \Pi(V_\mathrm{M}\leq 0 \mid Y),\quad
    \pi_\mathrm{W} = \Pi(V_\mathrm{W}\leq 0 \mid Y),\quad
    \pi_\mathrm{I} = 1 - \pi_\mathrm{W}.
\end{equation}
Since $\mathrm{SST} \Rightarrow \mathrm{MST} \Rightarrow \mathrm{WST}$, the corresponding exclusive class probabilities can be obtained by differencing.
Under Definition \ref{def:ST}, WST is a stochastically transitive condition, so the proposed framework yields direct posterior evidence for whether the latent comparison structure lies in the stochastically transitive regime or the SIT regime.

Furthermore, exact LST is equivalent to $\max_{\{i,j,k\}\in\mathcal{T}} |\mathcal{C}_{ijk}| = 0$, and this event has posterior probability zero under a continuous prior.
Approximate LST may instead be assessed through a practically negligible tolerance $\delta > 0$.
For the ranking problem considered here, however, the primary distinction is whether WST holds, under which the latent comparison probabilities induce a complete preorder.
This motivates the decision summaries under SIT introduced next.

\subsection{Decision Summaries under Intransitivity}
\label{subsec:decision_summaries}
We develop two complementary decision summaries from the posterior distribution of the latent match-up.
The blockwise ranking reports how finely the entities can be ordered, whereas the dominance graph retains individually credible pairwise directions that need not admit an ordering.
As established in Section~\ref{subsec:pairwise_comparisons}, a complete preorder compatible with all pairwise directions is available only under WST; under SIT, forcing a total order may obscure the intransitive structure.
We therefore seek summaries that separate entities only when the pairwise evidence supports such a separation, and otherwise withhold the comparison.

An ordered partition, which we call a blockwise ranking, provides such a summary.
It gives a coarse ranking in which between-block comparisons are established, whereas within-block comparisons are withheld.
Entities in the same block need be neither tied nor connected; rather, the block records only that the posterior preference relation introduced below does not separate them even when individual directions among them are credible.

\begin{df}[\textit{Blockwise Ranking}]
\label{def:blockwise-ranking}
\textup{
Let $R$ be an asymmetric binary relation on $\mathcal{V}$.
An ordered partition $\mathcal{B}(R) = (B_1,\ldots,B_g)$ of $\mathcal{V}$ is a \emph{blockwise ranking} with respect to $R$ if $i\mathrel{R}j$ whenever $i\in B_s$, $j\in B_t$, and $s<t$.
}
\end{df}

\begin{prp}[\textit{Existence and Uniqueness of the Finest Blockwise Ranking}]
\label{prp:uniqueness}
\textup{
Let $R$ be an asymmetric binary relation on a finite set $\mathcal{V}$.
Then there exists a unique blockwise ranking with respect to $R$ having the largest possible number of blocks, which we call the finest blockwise ranking.
}
\end{prp}

When WST holds, the \emph{Copeland score} \citep{copeland1951Reasonable, zoghi2015Copeland}, $c_i=\sum_{j\neq i}\mathbbm{1}\{p_{ij}>1/2\}$ for $i\in\mathcal{V}$, provides a directional score compatible with the complete preorder.
Since $p_{ij}>1/2$ is equivalent to $M_{ij}>0$, its Bayesian posterior expectation is $\mathbb{E}(c_i\mid Y) = \sum_{j\neq i} \Pi(M_{ij}>0 \mid Y)$.
Such scalar summaries are natural under stochastic transitivity, but under SIT they may force a linear ordering even when no complete preorder is compatible with the pairwise preferences.

Motivated by the Copeland principle, we instead construct a selective posterior preference relation.
Let $q_{ij}=\Pi(M_{ij}>0\mid Y)$ denote the posterior probability that $i$ beats $j$.
For a credibility threshold $\epsilon\in[1/2,1)$, we define the \emph{posterior preference relation} by $i\to_\epsilon j \ \Leftrightarrow \ q_{ij}>\epsilon$.
Because the posterior distribution assigns zero probability to $M_{ij}=0$, we have $q_{ij}+q_{ji}=1$.
Hence, $i\to_\epsilon j$ implies $q_{ji}\leq 1-q_{ij}<1-\epsilon\leq\epsilon$, so $j\to_\epsilon i$ cannot hold.
Thus, $\to_\epsilon$ is asymmetric, and Proposition~\ref{prp:uniqueness} guarantees the existence and uniqueness of the finest blockwise ranking $\mathcal{B}(\to_\epsilon)$.
In this sense, $\to_\epsilon$ keeps the directional rule underlying the Copeland score, but declares a direction only when its posterior credibility $q_{ij}$ exceeds the threshold $\epsilon$, withholding pairs whose probability lies near $1/2$.

Using the same posterior preference relation, define the dominance graph
\begin{equation}
\mathcal{D}_\epsilon = (\mathcal{V}, \mathcal{A}_\epsilon),\quad
\mathcal{A}_\epsilon = \{(i,j)\in\mathcal{V}^2 \mid i\to_\epsilon j\}.
\end{equation}
Because $\to_\epsilon$ is asymmetric, the graph contains at most one directed arc for each unordered pair.
Unlike the blockwise ranking, however, the dominance graph need not be acyclic or transitive and may therefore retain directed cycles supported by the posterior evidence.

The two summaries serve complementary purposes.
The finest blockwise ranking reports how finely the entities can be ordered while requiring every between-block direction to be credible, whereas the dominance graph displays the credible pairwise directions themselves, including directions that cannot be represented simultaneously by an ordered partition.
It is therefore particularly informative when the finest blockwise ranking is coarse or consists of a single block.

The threshold $\epsilon$ governs the trade-off between resolution and credibility.
To account for the multiplicity of simultaneous directional claims, we calibrate both summaries at a target BFDR level $\alpha$, using the posterior expected proportion of erroneous between-block claims for the blockwise ranking and of erroneous arcs for the dominance graph.
The formal definitions, threshold construction, and computational details are provided in Supplementary Material~S3.

\section{Simulation Study}
\label{sec:simulation}

\subsection{Simulation Design}
\label{subsec:simulation-design}
The CA-BIBT model is evaluated against four Bayesian competitors:
(i) the Bayesian Bradley--Terry (BBT) model, which imposes an LST structure as in the classical BT model;
(ii) the CARE model, implemented as a Bayesian curl-free submodel of the CA-BIBT model (see Section~\ref{subsec:related_work});
(iii) the BIBT model, the covariate-free version of the CA-BIBT model; and
(iv) the Intransitive Clustered Bradley--Terry (ICBT) model \citep{spearing2023Modeling}, which quantifies intransitivity as the discrepancy from a point estimate of the BT model via reversible-jump Markov chain Monte Carlo (RJMCMC).

Although the BBT and CARE models differ in their treatment of covariates, both are representative curl-free baselines and induce the same likelihood family up to reparameterization \citep[][Proposition~4]{dong2025Statistical}.
The ICBT model is fitted using their implementation, with the default hyperparameter values specified therein.
All models are run for $10{,}000$ MCMC iterations with the first $2{,}000$ discarded as burn-in; for the ICBT model, following the implementation, the burn-in comprises $1{,}000$ iterations of dimension estimation followed by $1{,}000$ iterations of standard posterior sampling.

Synthetic data are generated from the CA-BIBT model under a controlled setting.
Specifically, we consider a complete comparison graph with $N=10$ entities and $n_{ij}=20$ binary comparisons per edge.
Three edge-specific covariates are used ($d=3$), with the design matrix $X_E \in\mathbb{R}^{3\times 45}$ drawn from independent Gaussian entries to ensure full row rank; since $X_E$ is generated without imposing alignment with either flow space, the induced covariate flow generally contains both gradient and curl components.
For the CARE model, the projected matrix $X_{E,g} = X_E P_\mathrm{grad}$ is used in place of $X_E$.

The main simulation examines each model's robustness by varying the covariate contribution ratio $R_x$ over $\{0.1, 0.2, \ldots, 0.9\}$.
For each replication, $(\boldsymbol{u}, \boldsymbol{z}, \boldsymbol{\beta})$ in \eqref{eq:CA-BIBT_reparam} are drawn independently from standard normal distributions of the corresponding dimensions.
The resulting flows are then rescaled so that $\|\boldsymbol{M}\|_2^2 = |\mathcal{E}|$ and $R_{g,r} = R_{c,r} = (1-R_x)/2$.
Thus, the overall log-odds signal strength and the residual gradient--curl balance are fixed across all settings, isolating the effect of the gradient--curl composition of the covariate flow.

Performance is assessed along four indicators: mean squared error (MSE), sign-recovery accuracy, coverage probability, and execution time.
For an edge-flow component $\boldsymbol{A}$ and its posterior mean estimate $\widehat{\boldsymbol{A}}$, we define
\begin{equation}
    \mathrm{MSE}(\boldsymbol{A})
        = \frac{\|\widehat{\boldsymbol{A}}-\boldsymbol{A}\|_2^2} {|\mathcal{E}|},\quad
    \mathrm{Accuracy}(\boldsymbol{A})
        = \frac{\# \bigl\{\{i,j\}\in\mathcal{E} \mid \widehat{A}_{ij}A_{ij}>0\bigr\}} {|\mathcal{E}|}.
\end{equation}
The empirical coverage rate and average length of 95\% credible intervals are defined as $\sum_{\{i,j\}\in\mathcal{E}} \mathbbm{1}(A_{ij} \in \mathrm{CI}_{A_{ij}})/|\mathcal{E}|$ and $\sum_{\{i,j\}\in\mathcal{E}} |\mathrm{CI}_{A_{ij}}|/|\mathcal{E}|$, respectively, where $\mathrm{CI}_{A_{ij}}$ denotes the 95\% credible interval for $A_{ij}$.
The execution time is measured for the full MCMC run, including burn-in.
All measures are averaged over $100$ independent replications and reported for four quantities: the match-up vector $\boldsymbol{M}$, the total gradient flow $\boldsymbol{M}_g = \boldsymbol{M}_{g,r} + \boldsymbol{M}_{g,x}$, the total curl flow $\boldsymbol{M}_c = \boldsymbol{M}_{c,r} + \boldsymbol{M}_{c,x}$, and the covariate flow $\boldsymbol{M}_x$.

\subsection{Simulation Results}
\label{subsec:simulation-results}
\begin{figure}[htbp]
    \centering
    \includegraphics[width=1\linewidth]{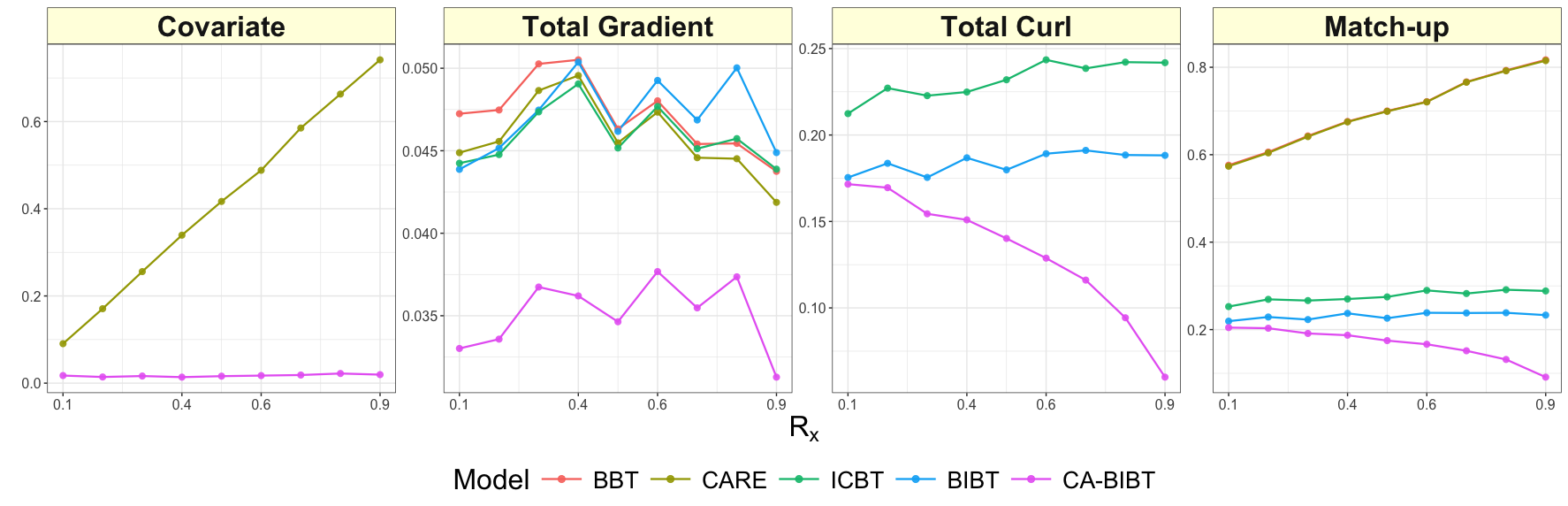}
    \caption{Mean squared error (MSE) as a function of $R_x$ for $N=10$ and $d=3$, shown for four components from left to right: the covariate flow $\boldsymbol{M}_x$, the total gradient flow $\boldsymbol{M}_g$, the total curl flow $\boldsymbol{M}_c$, and the match-up vector $\boldsymbol{M}$.}
    \label{fig:MSEs_N10}
\end{figure}

Curl-free models incur substantial misspecification error in match-up recovery.
The match-up panel of Figure~\ref{fig:MSEs_N10} shows that BBT and CARE attain markedly larger $\mathrm{MSE}(\boldsymbol{M})$ than the intransitive models (BIBT, CA-BIBT, and ICBT) across all values of $R_x$.
Among the intransitive models, the proposed Hodge-based models yield consistently smaller $\mathrm{MSE}(\boldsymbol{M})$ than ICBT, and CA-BIBT improves uniformly on BIBT with the gap widening as $R_x$ grows.
These results confirm that explicit curl representation is essential for accurate match-up recovery and that covariate information further reduces estimation error.

The component-wise MSE results show that covariates provide useful information for curl flow recovery.
The total curl panel of Figure~\ref{fig:MSEs_N10} reveals that CA-BIBT attains uniformly smaller $\mathrm{MSE}(\boldsymbol{M}_c)$ than BIBT and ICBT, with the gap widening rapidly as $R_x$ increases, since covariates aid identification in the high-dimensional curl flow space.
In the covariate panel, $\mathrm{MSE}(\boldsymbol{M}_x)$ for CARE increases monotonically with $R_x$ because $R_{c,x}$ grows proportionally with $R_x$, whereas CA-BIBT maintains stable and substantially smaller $\mathrm{MSE}(\boldsymbol{M}_x)$ throughout.

\begin{figure}[htbp]
    \centering
    \includegraphics[width=1\linewidth]{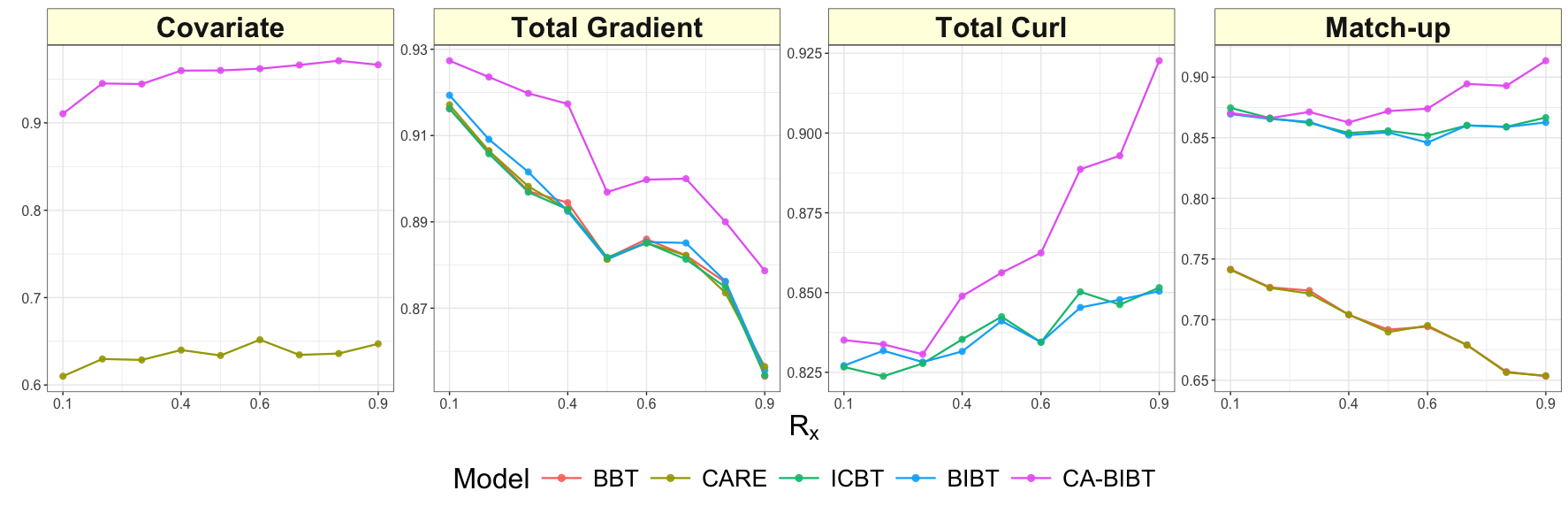}
    \caption{Sign-recovery accuracy as a function of $R_x$ for $N=10$ and $d=3$, shown for four components from left to right: the covariate flow $\boldsymbol{M}_x$, the total gradient flow $\boldsymbol{M}_g$, the total curl flow $\boldsymbol{M}_c$, and the match-up vector $\boldsymbol{M}$.}
    \label{fig:Accuracies_N10}
\end{figure}

As reported in Figure~\ref{fig:Accuracies_N10}, CA-BIBT achieves the highest sign-recovery accuracy across all settings.
The match-up panel shows that the accuracy of CA-BIBT improves steadily with $R_x$, whereas that of CARE declines monotonically, reflecting whether the growing covariate-induced curl component $R_{c,x}$ is captured by the model.
BIBT also maintains stably high accuracy throughout, indicating that explicit curl representation already accounts for much of the match-up sign structure under this design.
In the total curl panel, CA-BIBT progressively separates from BIBT and ICBT as $R_x$ increases, providing further evidence that covariate information facilitates sign recovery in the curl flow space.
The covariate panel indicates that the covariate flow accuracy of both CA-BIBT and CARE stabilizes once $R_x$ reaches a moderate level, with the remaining gap mainly reflecting their different representational capacities of the covariate flow.

As shown in Table~\ref{tab:N10}, the proposed models provide well-calibrated uncertainty quantification across all flow components, whereas curl-free models fail severely outside the gradient flow space.
Notably, CA-BIBT attains match-up coverage comparable to that of BIBT while producing shorter credible intervals.
This gain is concentrated in the curl flow, whose interval contracts markedly as $R_x$ increases (from $1.46$ to $0.92$) while the narrow gradient interval barely changes ($0.74$ to $0.71$), indicating that covariate information chiefly sharpens the high-dimensional curl component.
ICBT improves over the curl-free models for the match-up and total curl components, but its coverage remains substantially below that of the proposed models, and because it is reparameterized using the MLE of the BT model, its sampler does not produce posterior draws of the gradient flow.
Regarding computation, CA-BIBT and its submodels remain computationally efficient under the Gibbs sampling scheme, whereas ICBT incurs much higher computational cost because its RJMCMC scheme must explore both continuous parameters and discrete clustering structures.
Additional simulation results under alternative configurations are reported in Supplementary Material~S4.1.

\begin{table}[htbp]
    \centering
    \Large
    \caption{Average coverage probabilities and credible interval lengths of $95\%$ posterior credible intervals, and execution times across all values of $R_x$ for $N=10$.}
    \label{tab:N10}
    \resizebox{\textwidth}{!}{
    \begin{tabular}{cccccccccc}
        \toprule
        \multirow{3}{*}{Model} & \multicolumn{4}{c}{Coverage Probability}
        & \multicolumn{4}{c}{Credible Interval Length} & \multirow{3}{*}{\shortstack{Time\\(Seconds)}} \\
        \cmidrule(lr){2-5} \cmidrule(lr){6-9}
        & \multirow{2}{*}{Match-up} & Total & Total & Total & \multirow{2}{*}{Match-up} & Total & Total & Total \\
        & & Gradient & Curl & Covariate &  & Gradient & Curl & Covariate & \\
        \midrule
        BBT                 & 0.359 & 0.924 & --    & --    & 0.78 & 0.78 & --   & --   & 0.94 \\
        CARE                & 0.360 & 0.931 & --    & 0.283 & 0.78 & 0.78 & --   & 0.43 & 1.00 \\
        ICBT                & 0.669 & --    & 0.715 & --    & 1.01 & --   & 1.01 & --   & 639.43 \\
        \textbf{BIBT}       & 0.929 & 0.949 & 0.920 & --    & 1.74 & 0.85 & 1.52 & --   & 1.31 \\
        \textbf{CA-BIBT}    & 0.922 & 0.948 & 0.911 & 0.953 & 1.45 & 0.73 & 1.25 & 0.48 & 1.29 \\
        \bottomrule
    \end{tabular}
    }
\end{table}

\section{Empirical Applications}
\label{sec:applications}
We illustrate the proposed framework using two animal dominance datasets available through the R package \texttt{DomArchive} \citep{strauss2022DomArchive}.
Section~\ref{subsec:canary} demonstrates how the CA-BIBT model separates biologically interpretable covariate effects from residuals and illustrates the complementary roles of the finest blockwise ranking and the dominance graph under SIT.
Section~\ref{subsec:guanaco} focuses in particular on BFDR calibration for selecting an error-controlled finest blockwise ranking.
Together, the two applications illustrate how the proposed framework supports interpretable inference across the ST regimes.

\subsection{Application 1: Canary Dominance Data}
\label{subsec:canary}
The canary dominance data \citep{shoemaker1939Social} provide a natural setting in which to evaluate the CA-BIBT model, because prior observations documented cyclic dominance within this flock and biologically interpretable covariates are available.
The dataset consists of $n=10{,}693$ repeated decision fights among $N=10$ domestic canaries observed from June 1936 to March 1937.
The female birds are a14, a15, a17, a18, and a19, and the male birds are a39, a55, a58, a97, and a98; the mated pairs are (a14, a58), (a15, a55), (a17, a39), (a18, a98), (a19, a97).
\cite{shoemaker1939Social} documented two patterns: a prevailing dominance of males over females and a tendency for a female to dominate her own mate during the breeding season.
This setting allows us to examine the extent to which these documented attributes account for the aggregate comparison structure and whether a nontrivial curl component remains.

The covariate design incorporates two biologically motivated pair-specific covariates.
For each bird $i$, let $x_i=1$ if bird $i$ is male and $x_i=0$ otherwise, and let $b_{ij}=1$ if birds $i$ and $j$ are mates and $b_{ij}=0$ otherwise; for $i, j \in \mathcal{V}$, we define
\begin{equation}
    x_{\mathrm{sex},ij} = x_i - x_j,\quad
    x_{\mathrm{mate},ij} = b_{ij} (x_i - x_j),
\end{equation}
with corresponding coefficients $\beta_\mathrm{sex}$ and $\beta_\mathrm{mate}$.
The first covariate represents the overall directional sex effect, whereas the second captures a mate effect.
The sex covariate is a pure gradient flow, whereas the mate covariate carries components in both the gradient and curl flow spaces, and its gradient component reduces to $(2/N) \boldsymbol{x}_\mathrm{sex}$; hence, $d_s=d_c=1$.

Posterior inference under the CA-BIBT model supports aggregate counterparts of both documented patterns.
The posterior means of the sex effect $\beta_\mathrm{sex}$ and the mate effect $\beta_\mathrm{mate}$ are $1.77$ and $-1.94$, with 95\% credible intervals $[1.62,1.93]$ and $[-2.17,-1.72]$, respectively.
The positive sex effect reproduces the prevailing male dominance, whereas the negative mate effect is large enough to offset it.
For a mated pair oriented from the male to the female, the covariate flow contribution is $\mathbb{E}(\beta_\mathrm{sex}+\beta_\mathrm{mate}\mid Y)=-0.16$, with $\Pi(\beta_\mathrm{sex} + \beta_\mathrm{mate}<0 \mid Y) = 0.98$, which is consistent with the within-mate female dominance reported by \cite{shoemaker1939Social}.
Hence, there is strong posterior evidence that mating reverses the general male advantage at the covariate level, although the total match-up direction also depends on the residuals.

The posterior mean contribution ratio is $\mathbb{E}(R_x\mid Y)=0.61$.
The total gradient and curl shares are $\mathbb{E}(R_g\mid Y)=0.57$ and $\mathbb{E}(R_c\mid Y)=0.43$, respectively.
Within these subspaces, the covariate shares are $\mathbb{E}(R_{x\mid g}\mid Y)=0.82$ and $\mathbb{E}(R_{x\mid c}\mid Y)=0.34$.
Thus, the observed attributes explain most of the hierarchy but leave most cycle-induced dominance to the residual curl term, which lies in a higher-dimensional space than the gradient term and is regularized by the horseshoe prior.

\begin{figure}[htbp]
    \centering
    \includegraphics[width=0.64\linewidth]{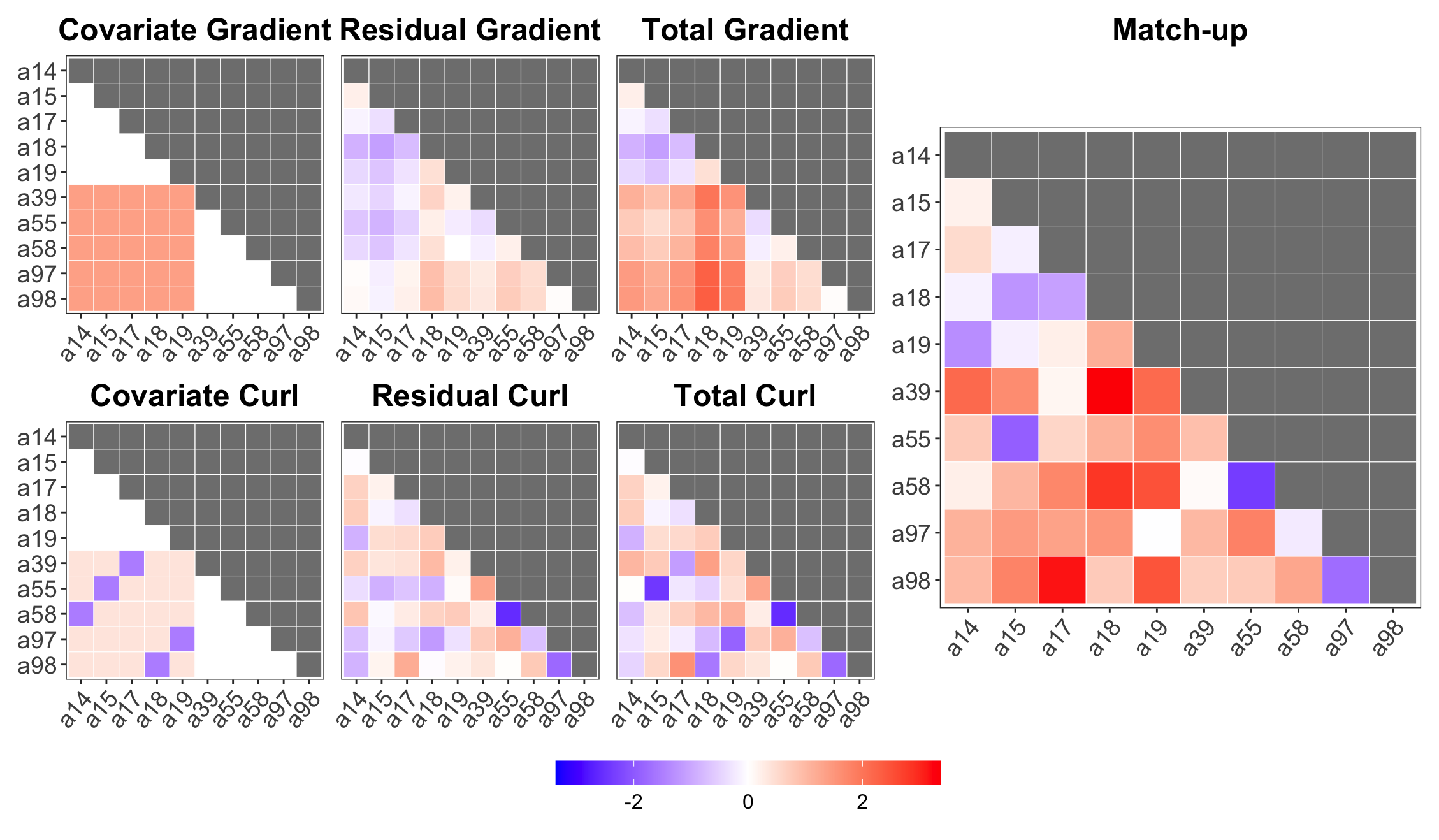}
    \includegraphics[width=0.34\linewidth]{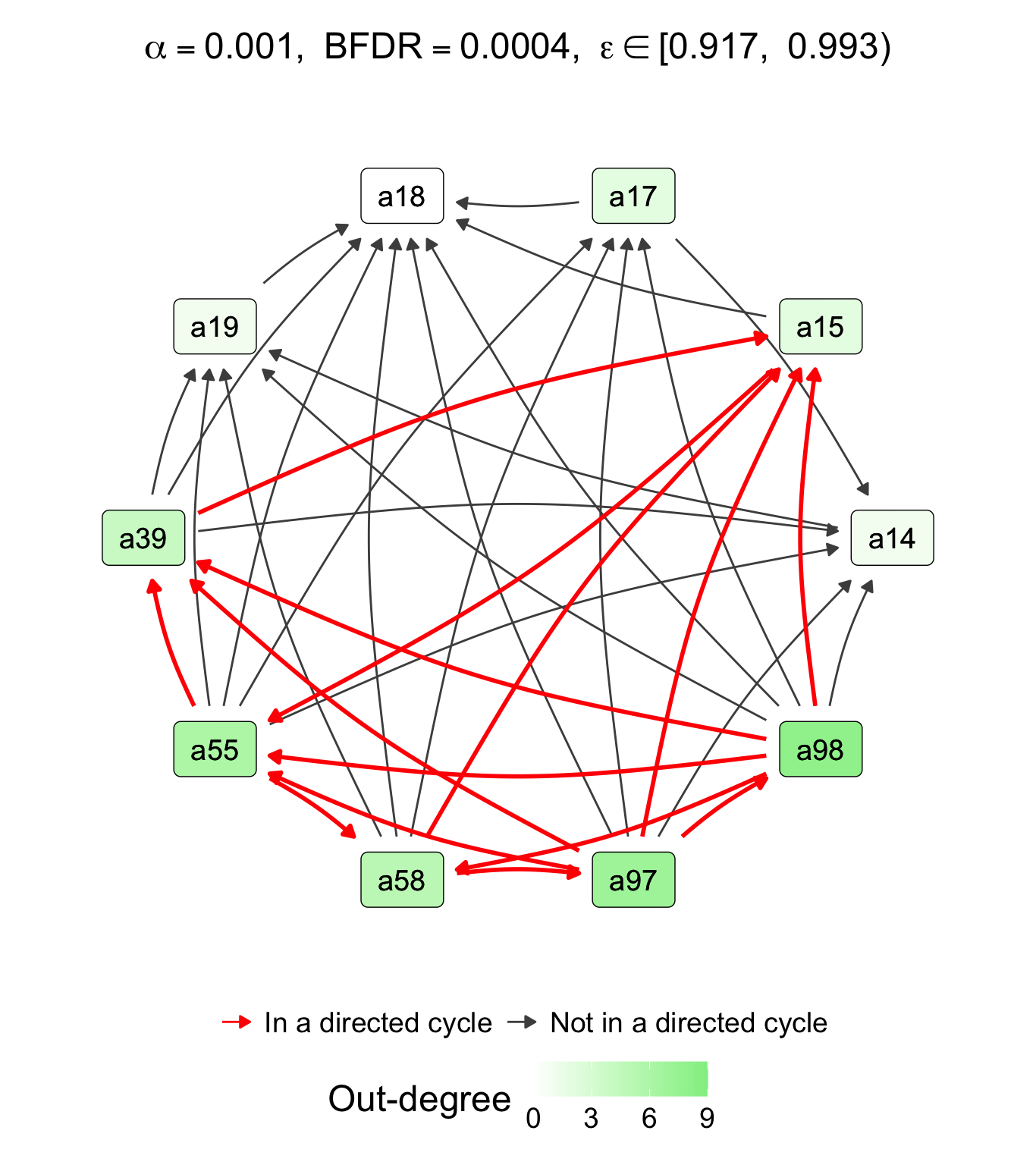}    
    \caption{
    Posterior summaries for the canary dominance data under the CA-BIBT model.
    Left: the posterior mean decomposition of pairwise effects, where the upper and lower panels show the covariate-induced, residual, and total gradient and curl flows, respectively, and the rightmost panel shows the total match-up values.
    Right: the BFDR-calibrated dominance graph at the target level $\alpha=0.001$.
    Positive entries indicate an advantage of the row bird over the column bird.
    }
    \label{fig:Flows_Canary}
\end{figure}

The posterior mean heatmaps in the left panel of Figure~\ref{fig:Flows_Canary} localize this attribution to individual pairs.
The covariate-induced gradient flow is a uniform male-over-female offset that vanishes within each sex, whereas the covariate-induced curl flow is negative on exactly the five mated pairs.
The residual curl flow, by contrast, is diffuse and sign-varying across the flock, including among birds of the same sex, and accounts for the unexplained majority of the cycle-induced structure.

All retained posterior draws fell in the SIT class, giving $\pi_\mathrm{I}=1$.
At the target level $\alpha=0.001$, the finest blockwise ranking consists of a single block and therefore yields no ordering.
At the same target level, the dominance graph retains several credible pairwise directions, shown in the right panel of Figure~\ref{fig:Flows_Canary}, and provides finer pairwise information than the blockwise ranking.
Every male has a larger out-degree (the number of outgoing arcs in the BFDR-calibrated dominance graph) than every female, and the graph contains, for example, the directed cycle a15 $\to$ a55 $\to$ a39 $\to$ a15.
The first between-block separation appears at $\alpha=0.013$ and remains the only separation for $\alpha>0.013$, placing a18 below the remaining birds.

The CARE comparison exposes a structural limitation of curl-free attribution.
With the sex attribute as the sole covariate, the CARE model assigns a larger overall covariate contribution than the CA-BIBT model, $\mathbb{E}(R_x\mid Y)=0.60$ versus $0.46$, but a smaller contribution within the gradient flow space, $\mathbb{E}(R_{x\mid g}\mid Y)=0.60$ versus $0.81$.
Because a curl-free fit collapses these two contributions into the same quantity, it cannot distinguish the overall role of a covariate from its role within the hierarchical structure.
The orthogonal Hodge decomposition makes this distinction identifiable and enables separate quantitative attribution of the hierarchical and cycle-induced dominance structure to observed attributes.

\subsection{Application 2: Guanaco Dominance Data}
\label{subsec:guanaco}
The guanaco dominance dataset \citep{correa2013Social} records $n=503$ agonistic interactions among $N=9$ adult individuals from a family group of guanacos (\textit{Lama guanicoe}), consisting of a single adult male (Male), two non-pregnant females (N.4 and N.8), and six pregnant females.
The posterior probability of WST is $\pi_\mathrm{W} = 0.63$, indicating that a total preorder compatible with all pairwise dominance relationships in this group is plausible but not decisively supported.
Because the blockwise ranking procedure depends only on the estimable match-up values $M_{ij}$, it is applicable to any paired comparison model.
For this application, we calibrate the credibility threshold using the BFDR-based strategy detailed in Supplementary Material~S3.1.

As displayed in Figure~\ref{fig:FBRs}, Male is identified as the most subordinate individual at every calibration level considered ($\epsilon < 0.875$), in agreement with the findings reported by \cite{correa2013Social}.
By contrast, the finer separations within the middle block of females are declared at thresholds close to $\epsilon = 1/2$, where the corresponding directional claims are supported by little posterior evidence.
The sensitivity analysis in Supplementary Material~S4.3 shows that the high-credibility separations obtained under stringent BFDR calibration are stable across the residual curl shrinkage priors, although the corresponding ranges of $\epsilon$ shift across priors.
However, the finer distinctions that emerge near $\epsilon=1/2$ are sensitive to the prior specification.
By adopting the posterior preference relation $\rightarrow_\epsilon$ with BFDR calibration at a target level $\alpha$, practitioners obtain a principled correspondence between a tolerable false discovery rate and a decision summary that remains valid under any ST class.

\begin{figure}[htbp]
    \centering
    \includegraphics[width=1\linewidth]{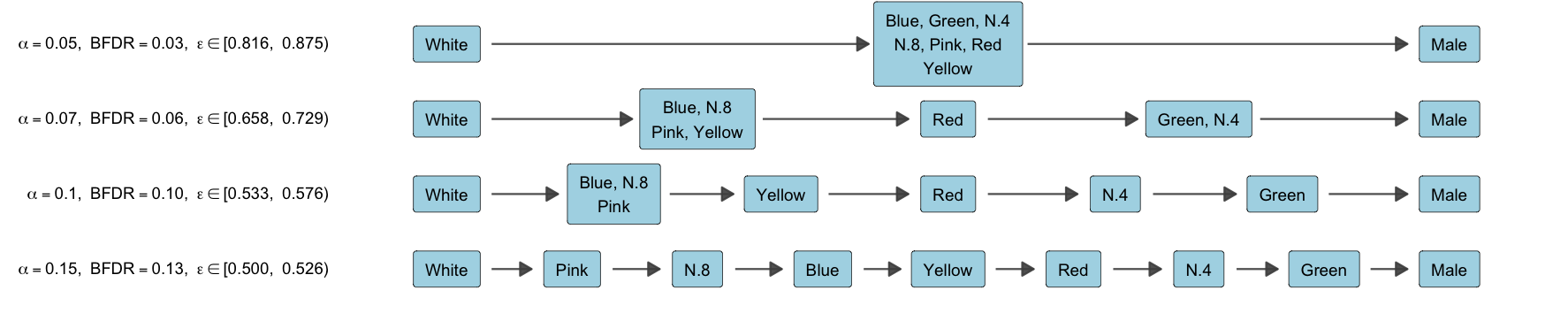}
    \caption{
    BFDR-calibrated finest blockwise rankings for the adult guanaco dominance data, where each row corresponds to the finest blockwise ranking whose BFDR does not exceed the target level $\alpha$.
    Nodes represent individuals or blocks of individuals, and arrows indicate between-block dominance declarations.
    The displayed interval gives the range of credibility thresholds $\epsilon$ that produce the same finest blockwise ranking.
    }
    \label{fig:FBRs}
\end{figure}

\section{Discussion}
\label{sec:discussion}
The central contribution of the proposed framework is to make the role of covariates in intransitive comparisons explicit and quantifiable.
By embedding a combinatorial Hodge decomposition into a logistic model for binary comparisons, the CA-BIBT model separates the latent match-up into mutually orthogonal covariate, residual gradient, and residual curl flows under explicit identifiability constraints.
This directly addresses the question raised in the introduction---how far observed attributes account for the intransitivity in the data---by attributing to covariates whatever structure they can explain while quantifying the cycle-induced structure that they cannot.
Posterior inference for the latent match-up further supports two complementary BFDR-calibrated decision summaries: a blockwise ranking that reports how finely the entities can be ordered and a dominance graph that retains credible pairwise directions, including cyclic relations.

A primary limitation is that the CA-BIBT model accommodates only static edge-specific covariates, not comparison-specific covariates that vary across individual contests between the same pair, such as home advantage or weather conditions in sports \citep[see][]{agresti2012Categorical, caron2012Efficient}.
The canary analysis illustrates why this extension matters: \cite{shoemaker1939Social} observed that male--female dominance could reverse during the breeding season.
Such reversals need not eliminate a static effect in the aggregated data, but the present analysis cannot identify when or how the effect changes over time, and the resulting unmodeled temporal heterogeneity may be reflected in both the residual gradient and residual curl flows.
Incorporating such covariates, however, introduces multiple edges between the same pair, so that the underlying graph is no longer simple and the combinatorial Hodge decomposition \citep{jiang2011Statistical} does not apply directly; preserving a Hodge-theoretic perspective in this setting is a compelling direction for future research.

A further limitation concerns scalability: the residual curl update requires Cholesky factorization of a $q_z \times q_z$ precision matrix with $q_z = O(N^2)$, so the computational cost grows rapidly with the number of entities $N$.
One way to avoid this bottleneck is the residual-curl-free submodel of Supplementary Material~S2, which is effective only when the covariate design adequately spans the high-dimensional curl flow space, but is of limited use in general.
More broadly applicable alternatives are scalable posterior approximations such as variational Bayes \citep{blei2017Variational, durante2019Conditionally}.

Despite these limitations, the proposed framework offers, to our knowledge, the first treatment of binary paired comparisons that quantifies covariate-induced and residual cycle-induced structure within a single identifiable model, turning the qualitative observation that covariates may shape intransitivity into a quantity that can be estimated with calibrated uncertainty.

\section*{Acknowledgement}
This work was supported by JST SPRING (Grant Number JPMJSP2151) and by the Japan Society for the Promotion of Science (JSPS) KAKENHI (Grant Numbers 23K13019, 24K21420 and 25H00546).


\putbib[refs-BIBT]
\end{bibunit}

\newpage
\setcounter{equation}{0}
\setcounter{section}{0}
\setcounter{table}{0}
\setcounter{figure}{0}
\setcounter{page}{1}
\setcounter{df}{0}
\setcounter{thm}{0}
\setcounter{lem}{0}
\setcounter{cor}{0}
\setcounter{prp}{0}
\setcounter{exm}{0}
\setcounter{algo}{0}
\setcounter{as}{0}
\setcounter{rem}{0}

\renewcommand{\thesection}{S\arabic{section}}
\renewcommand{\theequation}{S\arabic{equation}}
\renewcommand{\thetable}{S\arabic{table}}
\renewcommand{\thefigure}{S\arabic{figure}}
\renewcommand{\thedf}{S\arabic{df}}
\renewcommand{\thethm}{S\arabic{thm}}
\renewcommand{\thelem}{S\arabic{lem}}
\renewcommand{\thecor}{S\arabic{cor}}
\renewcommand{\theprp}{S\arabic{prp}}
\renewcommand{\theexm}{S\arabic{exm}}
\renewcommand{\thealgo}{S\arabic{algo}}
\renewcommand{\theas}{S\arabic{as}}
\renewcommand{\therem}{S\arabic{rem}}

\vspace{1cm}
\begin{center}
{\LARGE
{\bf Supplementary Material for ``The Covariate-Assisted Bayesian Intransitive Bradley--Terry Model via Combinatorial Hodge Theory"}
}
\end{center}

This Supplementary Material details the BIBT specialization of the CA-BIBT model, model reductions and connections to existing frameworks, and the BFDR calibration procedure.
It also presents additional numerical studies and provides proofs of the theoretical results stated in the main article and this supplement.

\begin{bibunit}[chicago]

\section{The Bayesian Intransitive Bradley--Terry Model}
\label{supplementary:BIBT}
When edge-specific covariate information is unavailable, the covariate flow $X_E^\top\boldsymbol{\beta}$ cannot be specified, and the latent match-up is modeled through its gradient and curl components alone.
We refer to this covariate-free specialization of the CA-BIBT model as the \emph{Bayesian Intransitive Bradley--Terry} (BIBT) model.
The BIBT model serves as the intransitive but covariate-free baseline in the simulation study of Section~5 and underlies the guanaco dominance analysis of Section~6.2.
Formally, the BIBT model assumes that the latent match-up function $M\in L^2_\wedge(\mathcal{E})$ admits the decomposition $M=\mathrm{grad}\, s+\mathrm{curl}^\ast \Phi$ with $s\in L^2(\mathcal{V})$ and $\Phi\in L^2_\wedge(\mathcal{T})$, or, in coordinate form, 
\begin{equation}
\label{def:BIBT}
    \boldsymbol{M} = G\boldsymbol{s} + C^\top\boldsymbol{\Phi},
\end{equation}
with $\boldsymbol{s} \in \mathbb{R}^N$ and $\boldsymbol{\Phi} \in\mathbb{R}^{|\mathcal{T}|}$.
Equivalently, for each pair $(i,j)$,
\begin{equation}
\label{def:match-up}
    M_{ij} = s_i-s_j + \sum_{k:\{i,j,k\}\in\mathcal{T}}\Phi_{ijk}.
\end{equation}
The first term in \eqref{def:BIBT} is the gradient flow induced by pairwise differences between the score parameters, and the second term is the curl flow, encoding cycle-induced structure.
The BIBT model is the probabilistic model $p_{ij}=\sigma(M_{ij})$ with $M_{ij}$ as in \eqref{def:match-up}.

Because the BIBT model omits the covariate flow, the orthogonality constraints that separate the residual flows from the covariate flow in the CA-BIBT model are no longer needed, and the only sources of non-identifiability are the kernels of the two operators.
Since $\ker(\mathrm{grad})=\mathrm{span}(\boldsymbol{1})$, the score vector $\boldsymbol{s}$ is identifiable only up to an additive constant; likewise, since $\mathrm{curl}^\ast$ has a nontrivial kernel, the triangular parameter $\boldsymbol{\Phi}$ is identifiable only up to elements of $\ker(\mathrm{curl}^\ast)$.
Let $K=\dim\,\mathrm{im}(\mathrm{curl}^\ast) = \binom{N-1}{2}$, and let $A\in\mathbb{R}^{|\mathcal{T}|\times(|\mathcal{T}|-K)}$ be a full-column-rank matrix whose columns form a basis for $\ker(\mathrm{curl}^\ast)$.
To remove the two indeterminacies, we impose $\boldsymbol{v}^\top\boldsymbol{s}=0$ and $V^\top\boldsymbol{\Phi}=\boldsymbol{0}$, where $\boldsymbol{v}\in\mathbb{R}^N$ satisfies $\boldsymbol{v}^\top\boldsymbol{1}\neq0$ and $V\in\mathbb{R}^{|\mathcal{T}|\times(|\mathcal{T}|-K)}$ satisfies $\det(V^\top A)\neq0$.
The following result, a direct specialization of Theorem 1, formalizes identifiability under these constraints.
\begin{thm}[\textit{Identifiability of the BIBT model}]
\label{thm:identifiability_BIBT}
\textup{
Let 
\begin{equation}
    \boldsymbol{M} = G \boldsymbol{s} + C^\top \boldsymbol{\Phi}
\end{equation}
with $\boldsymbol{s} \in \mathbb{R}^N$ and $\boldsymbol{\Phi} \in \mathbb{R}^{|\mathcal{T}|}$. 
Suppose that $\boldsymbol{v}^\top \boldsymbol{1}\neq 0$ and $\det(V^\top A)\neq 0$.
Define the constrained parameter space
\begin{equation}
    \Theta(\boldsymbol{v}, V) =
    \left\{ (\boldsymbol{s}, \boldsymbol{\Phi}) \ \middle| \ \boldsymbol{v}^\top \boldsymbol{s}=0,\ V^\top \boldsymbol{\Phi}=0 \right\}.
\end{equation}
Then the BIBT model is identifiable over $\Theta(\boldsymbol{v}, V)$; equivalently, the mapping $(\boldsymbol{s}, \boldsymbol{\Phi}) \in \Theta(\boldsymbol{v}, V) \mapsto \{ \sigma(M_{ij}) \}_{\{i,j\}\in\mathcal{E}}$ is injective.
}
\end{thm}

\noindent\textit{Proof.} The proof is given in Section~\ref{supplementary:proofs}.

The proof is the covariate-free reduction of the proof of Theorem~1: with $X_E^\top\boldsymbol{\beta}$ absent, the Hodge projections have no covariate term to act on, and the constraints $X_E G\boldsymbol{s}=0$ and $X_E C^\top\boldsymbol{\Phi}=0$ become vacuous.
The prior specification and Gibbs sampler for the BIBT model follow from those of the CA-BIBT model in Sections~3.2--3.3 by deleting the coefficient $\boldsymbol{\beta}$ and its associated updates.
Under the reparameterization (6), the reduced dimensions are $q_u=N-1$ and $q_z=K=\binom{N-1}{2}$, corresponding to $d_s=d_c=0$ in the general expressions $q_u=N-1-d_s$ and $q_z=K-d_c$.

\section{Model Reductions and Connections to Existing Frameworks}
\label{supplementary:special-cases}
The CA-BIBT model represents the latent match-up vector as
\begin{equation}
\label{sup:CA-BIBT}
    \boldsymbol{M} = G \boldsymbol{s} + C^\top \boldsymbol{\Phi} + X_E^\top \boldsymbol{\beta},
\end{equation}
the sum of a residual gradient flow $G\boldsymbol{s}$, a residual curl flow $C^\top\boldsymbol{\Phi}$, and a covariate flow $X_E^\top\boldsymbol{\beta}$.
This section shows how several transitive and intransitive paired-comparison models arise as special cases under structural restrictions on these three components, complementing the summary in Table 1.

\textbf{BIBT and BT models}:
Setting $\boldsymbol{\beta}=\boldsymbol{0}$ removes the covariate flow and yields the BIBT model \eqref{def:BIBT}, $\boldsymbol{M}=G\boldsymbol{s}+C^\top\boldsymbol{\Phi}$.
Further setting $\boldsymbol{\Phi}=\boldsymbol{0}$ removes the curl component and recovers the classical BT model \citep{bradley1952Rank}, $\boldsymbol{M}=G\boldsymbol{s}$, in which $p_{ij}=\sigma(s_i-s_j)$; identifiability is then ensured by a single constraint $\boldsymbol{v}^\top\boldsymbol{s}=0$ with $\boldsymbol{v}^\top\boldsymbol{1}\neq0$.

\textbf{CARE model}:
The \emph{covariate-assisted ranking estimation} (CARE) model \citep{fan2024Uncertainty} replaces edge-specific covariates by entity-specific ones.
Let $X=(\boldsymbol{x}_1,\dots,\boldsymbol{x}_N)\in\mathbb{R}^{d\times N}$ be the entity-specific covariate matrix and assume that the augmented matrix $\bar{X}=(\boldsymbol{1},X^\top)\in\mathbb{R}^{N\times(d+1)}$ has full column rank; the induced edge-specific design $X_E=XG^\top$ is then of full row rank.
The match-up vector is 
\begin{equation}
\label{def:CARE}
    \boldsymbol{M}=G(\boldsymbol{s}+X^\top\boldsymbol{\beta}),
\end{equation}
recovered from \eqref{sup:CA-BIBT} by setting $\boldsymbol{\Phi}=\boldsymbol{0}$ and specializing the covariate design to $X_E=XG^\top$.
Under this specialization, the covariate flow $GX^\top\boldsymbol{\beta}$ lies entirely in the gradient flow space, so $\boldsymbol{M}$ is a pure gradient flow and the induced winning probabilities satisfy the LST condition.

The reduction also matches the identifiability constraints.
With $\boldsymbol{\Phi}=\boldsymbol{0}$, the constraints $V^\top\boldsymbol{\Phi}=0$ and $X_E C^\top\boldsymbol{\Phi}=0$ hold trivially, leaving $\boldsymbol{v}^\top\boldsymbol{s}=0$ and $X_E G\boldsymbol{s}=0$.
Using $X_E = XG^\top$ and $G^\top G=N I - \boldsymbol{1}\boldsymbol{1}^\top$ on the complete latent comparison graph, we obtain
\begin{equation}
    X_E G\boldsymbol{s} 
    = X G^\top G\boldsymbol{s}
    = X (N I - \boldsymbol{1}\boldsymbol{1}^\top) \boldsymbol{s}.
\end{equation}
Taking $\boldsymbol{v}=\boldsymbol{1}$ gives $\boldsymbol{1}^\top\boldsymbol{s}=0$, hence $X\boldsymbol{1}\boldsymbol{1}^\top\boldsymbol{s}=\boldsymbol{0}$ and $X_E G\boldsymbol{s}=N X\boldsymbol{s}$, so $X_E G\boldsymbol{s}=0$ is equivalent to $X\boldsymbol{s}=\boldsymbol{0}$.
The CA-BIBT constraints thus reduce to $\boldsymbol{1}^\top\boldsymbol{s}=0$ and $X\boldsymbol{s}=\boldsymbol{0}$.
These are exactly the centering and covariate-orthogonality conditions that separate the residual score $\boldsymbol{s}$ from the covariate effect $\boldsymbol{\beta}$: under the full-column-rank assumption on $\bar{X}$, they decompose the latent score $\boldsymbol{s}+X^\top\boldsymbol{\beta}$ uniquely into a covariate part $X^\top\boldsymbol{\beta}$ and an orthogonal residual $\boldsymbol{s}$, in agreement with the identifiability conditions of \cite{fan2024Uncertainty}.

\textbf{Structured Bradley--Terry model}:
The covariate-only structured BT model of \cite{springall1973Response}, listed as ``Structured BT'' in Table 1, is obtained from the CARE model by further removing the residual gradient flow.
Setting $\boldsymbol{s}=\boldsymbol{0}$ in the CARE specification \eqref{def:CARE} leaves $\boldsymbol{M}=GX^\top\boldsymbol{\beta}$, in which the latent match-up is generated entirely by the entity-specific covariates through the gradient operator.
Within the CA-BIBT model, this corresponds to $\boldsymbol{s}=\boldsymbol{0}$, $\boldsymbol{\Phi}=\boldsymbol{0}$, and $X_E=XG^\top$, so that the covariate flow remains gradient-confined and the induced winning probabilities again satisfy the LST condition.
Because no residual score or curl component is present, the identifying constraints on $\boldsymbol{s}$ and $\boldsymbol{\Phi}$ are unnecessary, and the coefficient $\boldsymbol{\beta}$ is identifiable whenever $\bar{X}=(\boldsymbol{1}, X^\top)$ has full column rank, the same condition assumed for the CARE model.

\textbf{Curl-free residual reduction}:
The reductions above are curl-free, retaining only transitive structure.
A complementary reduction retains cycle-induced structure while economizing on computation.
Setting $\boldsymbol{\Phi}=\boldsymbol{0}$ in \eqref{sup:CA-BIBT} but keeping a general edge-specific covariate design gives
\begin{equation}
    \boldsymbol{M} = G\boldsymbol{s} + X_E^\top\boldsymbol{\beta},
\end{equation}
a submodel of the CA-BIBT model that omits the residual curl flow $C^\top\boldsymbol{\Phi}$ while still admitting cyclic structure through the covariate flow.
Unlike the CARE and structured BT models, the design $X_E$ is not restricted to the entity-specific form $XG^\top$, so the covariate flow generally has a nonzero curl component $P_\mathrm{curl} X_E^\top\boldsymbol{\beta}$, where $P_\mathrm{curl}$ denotes the orthogonal projection in $L^2_\wedge(\mathcal{E})$ onto the curl flow space.
Consequently, the induced winning probabilities may violate the LST condition, and the model represents the intransitivity that the observed covariates can explain.

The motivation is computational: the residual curl update of Section~3.3 requires a Cholesky factorization of a $q_z\times q_z$ precision matrix with $q_z=K-d_c=O(N^2)$, which dominates the computational cost and becomes demanding for large $N$.
Removing $C^\top\boldsymbol{\Phi}$ eliminates this step while preserving covariate-driven cycle-induced structure.

Identifiability follows from Theorem~1 with $\boldsymbol{\Phi}=\boldsymbol{0}$: under $\boldsymbol{v}^\top\boldsymbol{1}\neq0$ and full-row-rank $X_E$, the constraints $\boldsymbol{v}^\top\boldsymbol{s}=0$ and $X_E G\boldsymbol{s}=0$ render $(\boldsymbol{s},\boldsymbol{\beta})$ identifiable.
This specification is best suited to settings in which the intransitivity is plausibly attributable to measured edge-specific covariates, since any residual curl structure not spanned by the covariate design is left unmodeled.

\section{Bayesian False Discovery Rate Calibration}
\label{supplementary:BFDR}
This section describes a \emph{Bayesian false discovery rate} (BFDR) calibration rule for the two decision summaries introduced in Section~4.2.
Following the Bayesian decision-theoretic formulation of \cite{muller2006FDR}, we evaluate the false discovery proportion conditional on the observed data by averaging over the posterior distribution of the latent truth indicators.
The goal is to control the posterior expected proportion of erroneous directional claims declared by each summary: the between-block claims of the blockwise ranking (Section~\ref{sup:BFDR-blockwise}) and the credible edges of the dominance graph (Section~\ref{sup:BFDR-graph}).
For background on FDR control, see \cite{benjamini1995Controlling, efron2001Empirical, storey2003Positive}.

\subsection{Calibration for the Blockwise Ranking}
\label{sup:BFDR-blockwise}
Recall that $q_{ij} = \Pi(M_{ij}>0 \mid Y)$ denotes the posterior probability that entity $i$ defeats entity $j$.
Throughout this section, we assume, as in Section~4.2, that the posterior distribution of each $M_{ij}$ is continuous, so that $q_{ij}+q_{ji}=1$ for all $i\neq j$.
For a credibility threshold $\epsilon\in[1/2,1)$, the posterior preference relation is defined as $i \to_\epsilon j \ \Leftrightarrow \ q_{ij}>\epsilon$.
Since $\epsilon\geq 1/2$, the relation $\to_\epsilon$ is asymmetric, and by Proposition~1 it induces a unique finest blockwise ranking, which we denote by $\mathcal{B}(\to_{\epsilon})$.
We collect the attainable finest blockwise rankings as $\mathfrak{B} = \left\{ \mathcal{B}(\to_\epsilon) \ \middle| \ \epsilon\in \bigl[\tfrac{1}{2},1 \bigr) \right\}$.

To make the correspondence with \cite{muller2006FDR} explicit, let the latent truth indicator be $r_{ij} = \mathbbm{1}\{M_{ij}>0\}$.
For a blockwise ranking $\mathcal{B}\in\mathfrak{B}$, define the set of between-block directional claims $\mathcal{P}(\mathcal{B}) = \left\{(i,j) \mid i\in B_s,\ j\in B_t,\ s<t \right\}$ and the decision indicator $\delta_{ij}(\mathcal{B}) = \mathbbm{1}\{(i,j)\in\mathcal{P}(\mathcal{B})\}$.
Thus, $\delta_{ij}(\mathcal{B})=1$ means that the reported blockwise ranking declares the directional claim $M_{ij}>0$.

A false directional discovery occurs when a declared claim is false.
The realized false discovery proportion (FDP) associated with $\mathcal{B}$ is therefore
\begin{equation}
    \mathrm{FDP}(\mathcal{B}) 
    = \frac{\sum_{i,j\in\mathcal{V}} (1-r_{ij}) \delta_{ij}(\mathcal{B})}
    {|\mathcal{P}(\mathcal{B})|\vee 1}
    = \frac{\sum_{(i,j)\in\mathcal{P}(\mathcal{B})} \mathbbm{1}\{M_{ij}\leq0\}}
    {|\mathcal{P}(\mathcal{B})|\vee 1}.
\end{equation}
The convention $|\mathcal{P}(\mathcal{B})|\vee 1 = \max\{|\mathcal{P}(\mathcal{B})|, 1\}$ avoids division by zero when the reported ranking makes no between-block directional claim.

The BFDR associated with $\mathcal{B}$ is the posterior expectation of this quantity.
Since $M_{ji}=-M_{ij}$ and $\Pi(M_{ij}=0\mid Y)=0$, the posterior probability of a directional error for the declared claim $(i,j)$ is $q_{ji} = \Pi(M_{ji}>0\mid Y)$, and hence
\begin{equation}
\label{eq:BFDR-blockwise}
    \mathrm{BFDR}(\mathcal{B})
    = \mathbb{E} \left[\mathrm{FDP}(\mathcal{B}) \mid Y \right]
    = \frac{\sum_{(i,j)\in\mathcal{P}(\mathcal{B})} q_{ji}}
    {|\mathcal{P}(\mathcal{B})|\vee 1}.
\end{equation}
When $|\mathcal{P}(\mathcal{B})|=0$, the blockwise ranking makes no between-block directional claim, and the convention in \eqref{eq:BFDR-blockwise} gives $\mathrm{BFDR}(\mathcal{B})=0$.
By construction of $\mathcal{P}(\mathcal{B})$, every declared claim $(i,j)$ satisfies $q_{ij}>\epsilon$, so its posterior error probability obeys $q_{ji}<1-\epsilon$; the threshold $\epsilon$ is thus a uniform lower bound on the credibility of all reported between-block directions, and \eqref{eq:BFDR-blockwise} averages these error probabilities.

Next we describe the threshold plateau structure used to compute the calibrated blockwise ranking.
Let
\begin{equation}
\label{eq:breakpoints}
    1=\theta_0>\theta_1>\cdots>\theta_L>\theta_{L+1}=\frac{1}{2}
\end{equation}
be the decreasing sequence obtained by ordering the distinct values in $\{q_{ij}\mid q_{ij}>1/2\}\cup\{1,1/2\}$.
Because the posterior preference relation is defined by the strict rule $q_{ij}>\epsilon$, the active directional relations on the interval $\epsilon\in[\theta_{k+1},\theta_k)$ are precisely those satisfying $q_{ij}\geq \theta_k$, for $k=0,\ldots,L$.
Hence, the induced finest blockwise ranking $\mathcal{B}(\to_\epsilon)$ is constant over each breakpoint interval $[\theta_{k+1},\theta_k)$.

The following definition fixes the BFDR-based selection rule.
The subsequent proposition shows that the selected blockwise ranking is well-defined.
\begin{df}[\textit{BFDR-calibrated Blockwise Ranking}]
\label{def:BFDR-calibrated}
\textup{
For a target level $\alpha\in(0,1/2)$, a blockwise ranking $\mathcal{B}\in\mathfrak{B}$ is called \emph{$\alpha$-admissible} if $\mathrm{BFDR}(\mathcal{B}) \leq \alpha$.
Let $\mathfrak{B}_\alpha = \{\mathcal{B}\in\mathfrak{B} \mid \mathrm{BFDR}(\mathcal{B}) \leq \alpha\}$ denote the set of $\alpha$-admissible blockwise rankings.
A \emph{BFDR-calibrated blockwise ranking} at level $\alpha$ is the unique maximizer $\mathcal{B}_\alpha = \argmax_{\mathcal{B}\in\mathfrak{B}_\alpha} |\mathcal{P}(\mathcal{B})|$.
}
\end{df}

\begin{prp}[\textit{Well-definedness of the Blockwise-Ranking Calibration}]
\label{prp:BFDR-blockwise}
\textup{
For every $\alpha\in(0,1/2)$, the admissible set $\mathfrak{B}_\alpha$ is nonempty and 
\begin{equation}
    \mathcal{B}_\alpha = \argmax_{\mathcal{B}\in\mathfrak{B}_\alpha} |\mathcal{P}(\mathcal{B})|
\end{equation}
is well-defined.
Moreover, for any $0<\alpha\leq\alpha'<1/2$, $\mathcal{P}(\mathcal{B}_\alpha) \subseteq \mathcal{P}(\mathcal{B}_{\alpha'})$.
}
\end{prp}

\noindent\textit{Proof.} The proof is given in Section~\ref{supplementary:proofs}.

For each $\mathcal{B}\in\mathfrak{B}$, define its credibility-threshold set by $I(\mathcal{B}) = \bigl\{\epsilon\in \bigl[\tfrac{1}{2},1\bigr) \mid \mathcal{B}(\to_\epsilon)=\mathcal{B}\bigr\}$.
By the breakpoint construction and the nestedness of attainable claim sets, this set is a half-open interval, which we write as $I(\mathcal{B})=[\underline{\epsilon}(\mathcal{B}),\overline{\epsilon}(\mathcal{B}))$. 
The collection $\{I(\mathcal{B})\}_{\mathcal{B}\in\mathfrak{B}}$ forms a disjoint partition of $[1/2,1)$.
Because the same blockwise ranking may persist across several consecutive breakpoint intervals, $I(\mathcal{B})$ is generally more informative than any single threshold endpoint.

Given a target level $\alpha\in(0,1/2)$, we report the BFDR-calibrated blockwise ranking together with its credibility-threshold interval 
\begin{equation}
    I_{\alpha}=I(\mathcal{B}_{\alpha}) = [\underline{\epsilon}_\alpha,\overline{\epsilon}_\alpha).
\end{equation}
This interval records the range of credibility thresholds over which the same finest blockwise ranking is induced.
A wider interval means that the selected blockwise ranking is invariant over a larger range of $\epsilon$, whereas a narrower interval means that a neighboring threshold regime is reached by a smaller perturbation of $\epsilon$.
Regardless of the width of $I_\alpha$, $\mathcal{B}_\alpha$ remains the BFDR-calibrated decision summary at level $\alpha$.
Moreover, by Proposition~\ref{prp:BFDR-blockwise}, increasing the tolerable BFDR level can only add between-block directional claims, so the calibrated blockwise ranking becomes weakly finer.

In posterior computation, $q_{ij}$ is replaced by its Monte Carlo estimate
\begin{equation}
\label{eq:qhat}
    \widehat q_{ij} = \frac{1}{S} \sum_{s=1}^{S} \mathbbm{1}\{M_{ij}^{(s)}>0\},
\end{equation}
where $M_{ij}^{(s)}$ denotes the $s$-th posterior draw of the latent match-up value.
Since $M_{ij}^{(s)}=-M_{ji}^{(s)}$ and, under the continuous posterior, $M_{ij}^{(s)}=0$ occurs with probability zero, the empirical probabilities satisfy $\widehat{q}_{ij} + \widehat{q}_{ji}=1$.
Therefore, the same argument applies to the empirical plug-in construction based on $\widehat{q}_{ij}$.
The empirical breakpoints, attainable blockwise rankings, BFDR values in \eqref{eq:BFDR-blockwise}, and credibility-threshold intervals are computed by replacing $q_{ij}$ with $\widehat{q}_{ij}$ throughout, yielding the empirical plug-in counterpart of the calibrated summary.

\subsection{Calibration for the Dominance Graph}
\label{sup:BFDR-graph}
The blockwise ranking reports only those credible directions that qualify as between-block claims, and it withholds a credible direction $i\to_\epsilon j$ whenever placing $i$ and $j$ in separate blocks is not supported by the remaining pairwise directions.
When cyclic dominance is pervasive, this may yield a coarse partition, in the extreme case a single block.
The \emph{dominance graph} provides a complementary summary by retaining the credible pairwise directions themselves.

For a credibility threshold $\epsilon\in[1/2,1)$, define the arc set $\mathcal{A}_\epsilon = \{(i,j)\in\mathcal{V}^2 \mid q_{ij}>\epsilon\}$ where $q_{ij}=\Pi(M_{ij}>0\mid Y)$, and let $\mathcal{D}_\epsilon = (\mathcal{V},\mathcal{A}_\epsilon)$ be the corresponding dominance graph.
Each arc $(i,j)\in\mathcal{A}_\epsilon$ declares the directional claim $M_{ij}>0$, with posterior error probability $q_{ji}=1-q_{ij}$.
Unlike the blockwise ranking, $\mathcal{D}_\epsilon$ may contain directed cycles and therefore displays the intransitive structure directly.

Let $\mathfrak{D} = \left\{ \mathcal{D}_\epsilon \ \middle|\ \epsilon\in \bigl[\tfrac{1}{2},1\bigr) \right\}$ be the collection of attainable dominance graphs.
For a graph $\mathcal{D}\in\mathfrak{D}$, write $\mathcal{A}(\mathcal{D})$ for its arc set.
The BFDR of $\mathcal{D}$ is defined as the posterior expected proportion of erroneous declared arcs: 
\begin{equation}
\label{eq:BFDR-graph}
    \mathrm{BFDR}(\mathcal{D}) = \frac{\sum_{(i,j)\in\mathcal{A}(\mathcal{D})} q_{ji}}{|\mathcal{A}(\mathcal{D})|\vee 1}.
\end{equation}
This convention gives $\mathrm{BFDR}(\mathcal{D})=0$ when $\mathcal{A}(\mathcal{D})=\emptyset$.

\begin{df}[\textit{BFDR-calibrated Dominance Graph}]
\label{def:BFDR-graph}
\textup{
For a target level $\alpha\in(0,1/2)$, a dominance graph $\mathcal{D}\in\mathfrak{D}$ is called \emph{$\alpha$-admissible} if $\mathrm{BFDR}(\mathcal{D})\leq\alpha$.
Let $\mathfrak{D}_\alpha = \left\{ \mathcal{D}\in\mathfrak{D} \mid \mathrm{BFDR}(\mathcal{D})\leq\alpha \right\}$ denote the set of $\alpha$-admissible dominance graphs.
A \emph{BFDR-calibrated dominance graph} at level $\alpha$ is the unique maximizer $\mathcal{D}_\alpha = \argmax_{\mathcal{D}\in\mathfrak{D}_\alpha} |\mathcal{A}(\mathcal{D})|$.
}
\end{df}

Since the graphs in $\mathfrak{D}$ change only at the breakpoints in \eqref{eq:breakpoints}, the collection $\mathfrak{D}$ is finite and the above maximization is over a finite set.
Equivalently, the calibrated graph is obtained by scanning the breakpoint-induced dominance graphs in decreasing order of posterior credibility and retaining the largest graph whose BFDR does not exceed $\alpha$.

Because newly admitted arcs have weakly smaller posterior credibility and hence weakly larger posterior error probability, the BFDR is nondecreasing along this path.
The calibrated dominance graph is therefore uniquely determined, and its arc set grows monotonically with the target level, $\mathcal{A}(\mathcal{D}_\alpha) \subseteq \mathcal{A}(\mathcal{D}_{\alpha'})$ for $0<\alpha\le\alpha'<1/2$.

The dominance graph should be interpreted together with the blockwise ranking.
The blockwise ranking answers how finely the entities can be ordered without contradicting the credible directions, whereas the dominance graph displays the credible pairwise directions themselves.
It is therefore most informative when the calibrated blockwise ranking is coarse: even a single-block ranking may correspond to a nonempty dominance graph whose arcs reveal the partial and possibly cyclic dominance structure.

\section{Additional Numerical Studies}
\label{supplementary:NumericalStudies}
This section assesses the robustness of the conclusions in Sections~5 and~6 from three perspectives, in each of which the baseline design is otherwise held fixed.
Section \ref{sup:simulation_N30} investigates the effect of increasing the number of entities to $N=30$ and the effect of weakening the edge signal.
Section \ref{sup:incomplete} relaxes the graph completeness and quantifies how recovery on observed and unobserved edges degrades under graph incompleteness.
Section~\ref{sup:sensitivity} embeds the horseshoe prior on the residual curl flow in a broader family of global--local shrinkage priors and assesses whether the substantive findings of Section~6 are robust to the particular shrinkage specification.

\subsection{Simulation Study under Larger-Graph and Weak-Signal Settings}
\label{sup:simulation_N30}
We further examine the performance of the proposed method in a larger comparison graph with $N=30$ entities.
Unless otherwise stated, the simulation design follows that in Section~5.
The complete comparison graph then contains $|\mathcal{E}|=\binom{30}{2}=435$ edges, and the dimension of the curl flow space increases to $\binom{29}{2}=406$.
We keep $d=3$ edge-specific covariates and vary the covariate contribution ratio $R_x$ over the same grid as in the main simulation.

To assess robustness to weaker comparison signals, we additionally vary the overall signal strength of the latent match-up vector.
Specifically, we scale $\boldsymbol{M}$ so that
\begin{equation}
    \|\boldsymbol{M}\|_2^2 = \alpha |\mathcal{E}|,
\end{equation}
and compare $\alpha=1$ with $\alpha=0.25$.
Reducing $\alpha$ shrinks the latent log-odds toward zero, thereby moving the winning probabilities closer to $1/2$.
Thus, the setting $\alpha=0.25$ represents a weak-signal regime in which both magnitude estimation and sign recovery are intrinsically more difficult.
Since the computational cost of fitting the ICBT model becomes prohibitive for $N=30$, we exclude the model from this supplementary experiment.
For each combination of $R_x$ and $\alpha$, the reported results are averaged over $100$ independent replications.

To compare estimation errors across different signal-strength settings, we report the signal-normalized mean squared error.
For an edge-flow component $\boldsymbol{A}$ and its posterior mean estimate $\widehat{\boldsymbol{A}}$, we define
\begin{equation}
    \mathrm{sMSE}(\boldsymbol{A}) = \frac{\|\widehat{\boldsymbol{A}}-\boldsymbol{A}\|_2^2}{\alpha|\mathcal{E}|},
\end{equation}
where $\|\boldsymbol{M}\|_2^2 = \alpha|\mathcal{E}|$ is the total signal energy of the latent match-up vector.
This normalization prevents the weak-signal setting from appearing artificially favorable merely because the absolute norm of the true latent flow is smaller.
Accordingly, sMSE measures estimation error relative to the overall signal scale rather than in absolute units.

\begin{figure}[htbp]
    \centering
    \begin{subfigure}[t]{\linewidth}
        \centering
        \includegraphics[width=\linewidth]{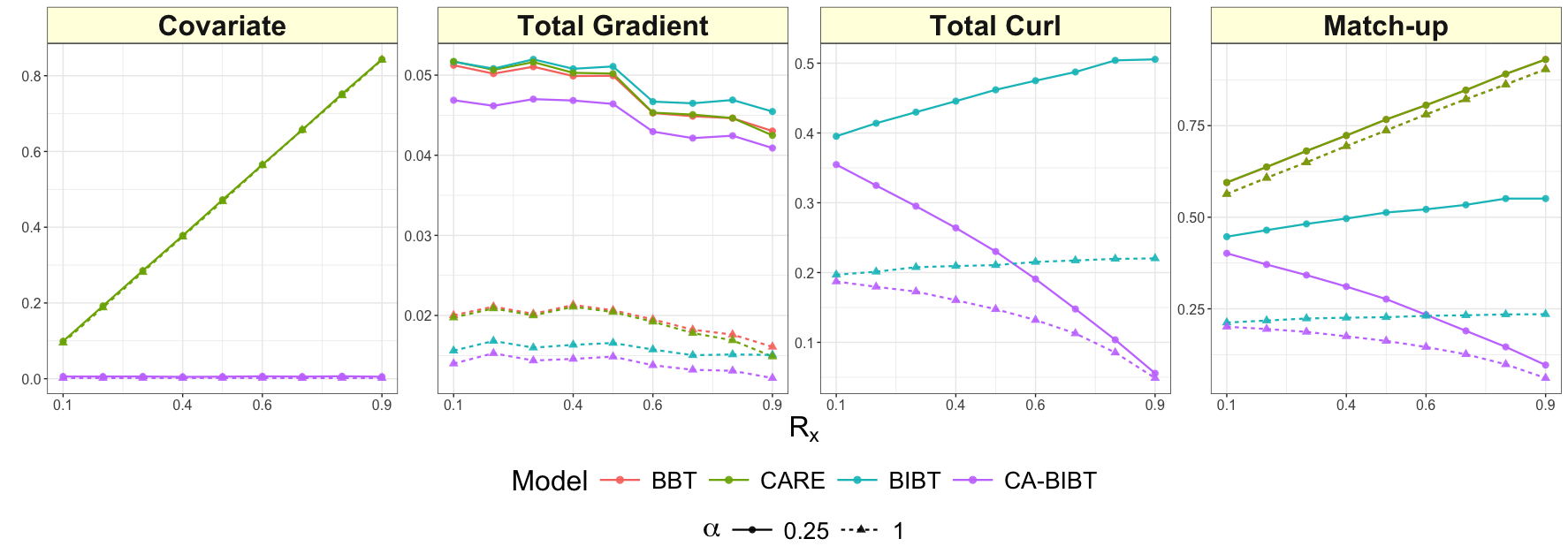}
        \caption{Signal-normalized MSE}
        \label{fig:N30_MSE}
    \end{subfigure}

    \vspace{0.5em}

    \begin{subfigure}[t]{\linewidth}
        \centering
        \includegraphics[width=\linewidth]{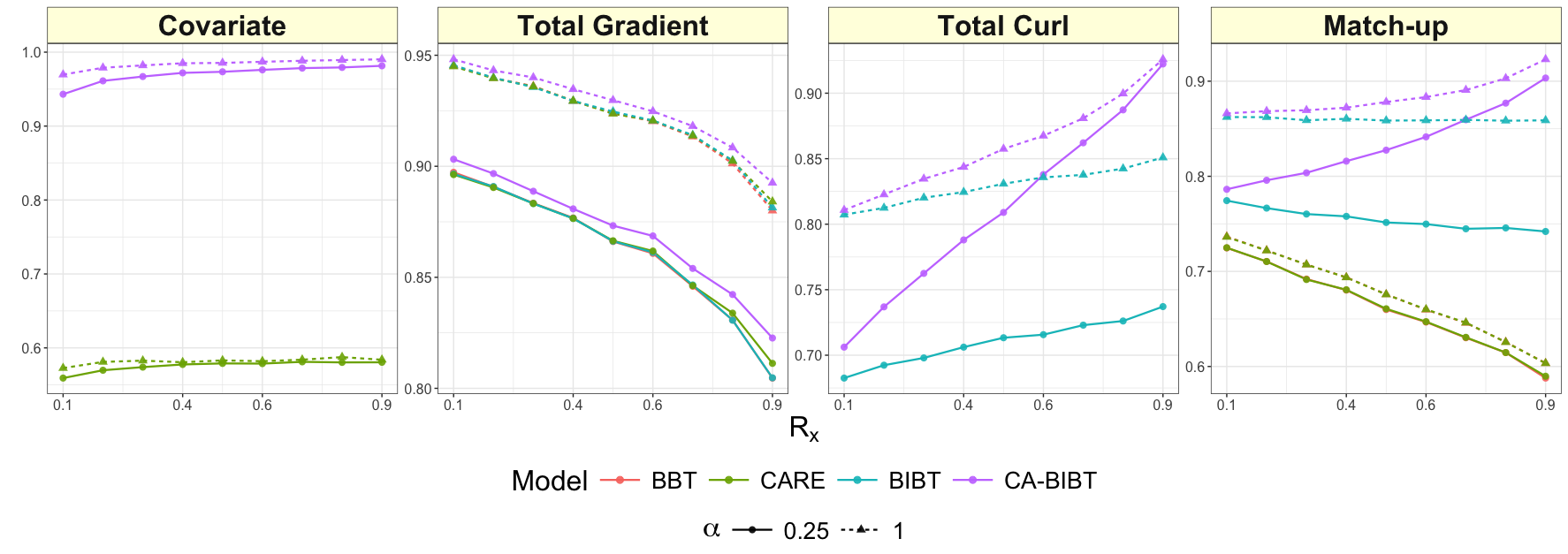}
        \caption{Sign-recovery accuracy}
        \label{fig:N30_accuracy}
    \end{subfigure}

    \caption{Estimation performance as a function of the covariate contribution ratio $R_x$ for $N=30$ and $d=3$ under two signal-strength settings, $\alpha=1$ and $\alpha=0.25$.
    Panel~(a) reports signal-normalized MSE, and panel~(b) reports sign-recovery accuracy.
    In each panel, the four component-specific plots correspond, from left to right, to the covariate flow $\boldsymbol{M}_x$, the total gradient flow $\boldsymbol{M}_g$, the total curl flow $\boldsymbol{M}_c$, and the match-up vector $\boldsymbol{M}$.}
    \label{fig:performance_N30}
\end{figure}

Figure~\ref{fig:performance_N30} summarizes magnitude recovery and sign recovery in panels~(a) and~(b), respectively.
The results show two main patterns.
First, CA-BIBT accurately recovers the covariate flow under both signal-strength settings.
The covariate-flow sMSE remains small, and the corresponding sign-recovery accuracy remains high even when $\alpha=0.25$.
This indicates that, when relevant edge-specific covariates are correctly specified, the covariate-induced component can still be identified reliably even when individual winning probabilities are close to $1/2$.

Second, the advantage of CA-BIBT over BIBT becomes more pronounced for the total curl flow and the match-up vector as $R_x$ increases.
This behavior is consistent with the simulation design.
Larger $R_x$ allocates a greater fraction of the latent match-up structure to the low-dimensional covariate flow space, while the residual gradient and residual curl components are reduced through $R_{g,r}=R_{c,r}=(1-R_x)/2$.
CA-BIBT can exploit this low-dimensional structure, whereas BIBT must represent the same covariate-induced signal indirectly through its latent gradient and curl components.
Consequently, CA-BIBT achieves smaller sMSE for the total curl flow and the match-up vector, and the same pattern is reflected in sign-recovery accuracy.

As expected, reducing $\alpha$ deteriorates sign-recovery performance, especially for the total curl flow and the match-up vector.
This deterioration reflects the fact that weaker latent match-up values make the signs of edge-level effects harder to infer.
Nevertheless, CA-BIBT maintains high covariate-flow accuracy and continues to improve total-curl and match-up recovery as $R_x$ increases.
These results suggest that the main advantage of CA-BIBT in the larger-graph setting is its ability to attribute weak comparison signals to the appropriate covariate-induced flow component.

Table~\ref{tab:N30} reports the average coverage probabilities defined as $\sum_{\{i,j\}\in\mathcal{E}} \mathbbm{1} (A_{ij} \in\mathrm{CI}_{A_{ij}}) / |\mathcal{E}|$, average credible interval lengths defined as $\sum_{\{i,j\}\in\mathcal{E}} |\mathrm{CI}_{A_{ij}}| / |\mathcal{E}|$, and execution times.
For the proposed models, the coverage probabilities for the total curl flow and the match-up vector improve as $\alpha$ increases, while the corresponding credible intervals also widen.
This indicates that stronger latent signals make the residual curl component more detectable; however, estimation remains uncertain because the curl flow space is high-dimensional.

A clearer distinction emerges when comparing BIBT and CA-BIBT.
CA-BIBT consistently yields shorter credible intervals than BIBT for the total curl flow and the match-up vector, as covariate information constrains exploration of the high-dimensional curl flow space.
This sharpening, however, carries a coverage cost: CA-BIBT undercovers both components relative to BIBT, while the low-dimensional gradient and covariate flows retain near-nominal coverage.

\begin{table}[htbp]
    \centering
    \Large
    \caption{Average coverage probabilities and credible interval lengths of $95\%$ posterior credible intervals, and execution times across all values of $R_x$ for $N=30$ under two signal-strength settings, $\alpha=1$ and $\alpha=0.25$.}
    \label{tab:N30}
    \resizebox{\textwidth}{!}{
    \begin{tabular}{ccccccccccc}
        \toprule
        \multirow{3}{*}{Model} & \multirow{3}{*}{$\alpha$} & \multicolumn{4}{c}{Coverage Probability}
        & \multicolumn{4}{c}{Credible Interval Length} & \multirow{3}{*}{\shortstack{Time\\(Seconds)}} \\
        \cmidrule(lr){3-6} \cmidrule(lr){7-10}
        & & \multirow{2}{*}{Match-up} & Total & Total & Total & \multirow{2}{*}{Match-up} & Total & Total & Total \\
        & &   & Gradient & Curl & Covariate &  & Gradient & Curl & Covariate & \\
        \midrule
        \multirow{2}{*}{BBT}
            & 0.25 & 0.386 & 0.954 & --    & --    & 0.44 & 0.44 & --   & --   & 15.2 \\
            & 1    & 0.212 & 0.896 & --    & --    & 0.45 & 0.45 & --   & --   & 15.7 \\        
        \multirow{2}{*}{CARE}
            & 0.25 & 0.388 & 0.954 & --    & 0.177 & 0.44 & 0.44 & --   & 0.13 & 13.9 \\
            & 1    & 0.212 & 0.909 & --    & 0.093 & 0.45 & 0.45 & --   & 0.14 & 13.5 \\
        \multirow{2}{*}{\textbf{BIBT}}
            & 0.25 & 0.882 & 0.954 & 0.869 & --    & 1.12 & 0.44 & 1.03 & --   & 91.9 \\
            & 1    & 0.919 & 0.949 & 0.916 & --    & 1.66 & 0.49 & 1.58 & --   & 90.8 \\
        \multirow{2}{*}{\textbf{CA-BIBT}}
            & 0.25 & 0.854 & 0.955 & 0.800 & 0.953 & 0.74 & 0.42 & 0.60 & 0.14 & 93.5 \\
            & 1    & 0.884 & 0.950 & 0.865 & 0.947 & 1.23 & 0.46 & 1.13 & 0.15 & 91.9 \\
        \bottomrule
    \end{tabular}
    }
\end{table}

\subsection{Robustness to Graph Incompleteness}
\label{sup:incomplete}
Remark 2 emphasizes that, when the observed comparison graph is incomplete, posterior inference for unobserved edges necessarily relies on model-based extrapolation rather than direct identification from the likelihood.
We therefore examine how MSE and sign-recovery accuracy of the proposed models change as the observed edge density varies.
We also examine whether incorporating a subset of the true covariates mitigates the effect of graph incompleteness on estimation accuracy.

We consider $N=20$ entities and define the complete reference edge set as $\mathcal{E}_\mathrm{com}=\{\{i,j\} \mid 1 \leq i<j \leq N\}$, so that $|\mathcal{E}_\mathrm{com}| = 190$.
Synthetic data are generated from the same flow-generation and rescaling scheme as in Section~5.
Specifically, we generate the residual gradient, residual curl, and covariate components from CA-BIBT, rescale the resulting latent match-up vector so that $\|\boldsymbol{M}\|_2^2 = |\mathcal{E}_\mathrm{com}|$, and fix $R_{g,r}=R_{c,r}=(1-R_x)/2$.

To examine robustness under partial covariate specification, we generate data using $d_\mathrm{true}=10$ edge-specific covariates, but fit the models using the first $d\in\{0,3,7,10\}$ covariates.
The case $d=10$ corresponds to an oracle covariate specification, whereas $d=3$ and $d=7$ represent partial covariate specifications.
The case $d=0$ corresponds to BIBT and serves as the covariate-free baseline.
We fix the covariate contribution ratio at $R_x=0.3$, so that the latent structure contains a non-negligible covariate-induced component while still retaining substantial residual gradient and curl components.
This design allows us to assess whether partial covariate information improves recovery of each flow component under graph incompleteness.

To generate incomplete comparison graphs, for each value of $\rho$, we sample $\rho|\mathcal{E}_\mathrm{com}|$ edges uniformly at random without replacement from $\mathcal{E}_\mathrm{com}$ and denote the resulting edge set by $\mathcal{E}$, so that $\rho = |\mathcal{E}| / |\mathcal{E}_\mathrm{com}|$.
We also define the missing edge set $\mathcal{E}_\mathrm{miss} = \mathcal{E}_\mathrm{com} \setminus \mathcal{E}$.
We consider $\rho\in\{0.5, 0.6,\ldots,1\}$ and condition on the resulting observed graph being connected.
For each incomplete dataset, we fit BIBT for $d=0$ and CA-BIBT for $d\in\{3,7,10\}$, while evaluating recovery on the full latent edge set $\mathcal{E}_\mathrm{com}$ and the missing edge set $\mathcal{E}_\mathrm{miss}$.
For each combination of $\rho$ and $d$, the reported results are averaged over $1{,}000$ independent replications.
Accordingly, the results should be interpreted as empirical evidence for model-based extrapolation, rather than as evidence of identifiability under incomplete observation.

\begin{figure}[htbp]
    \centering
    \begin{subfigure}[t]{0.8\linewidth}
        \centering
        \includegraphics[width=\linewidth]
        {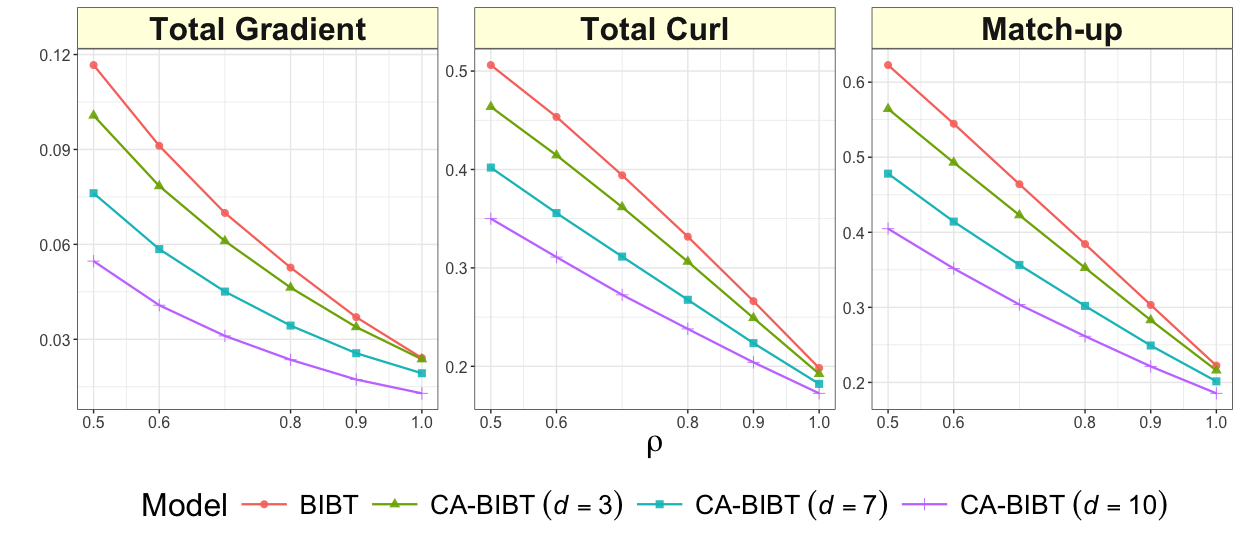}
        \caption{Evaluation on the complete reference edge set
        $\mathcal{E}_\mathrm{com}$.}
        \label{fig:MSEs_incom_complete}
    \end{subfigure}

    \vspace{0.5em}

    \begin{subfigure}[t]{0.8\linewidth}
        \centering
        \includegraphics[width=\linewidth]
        {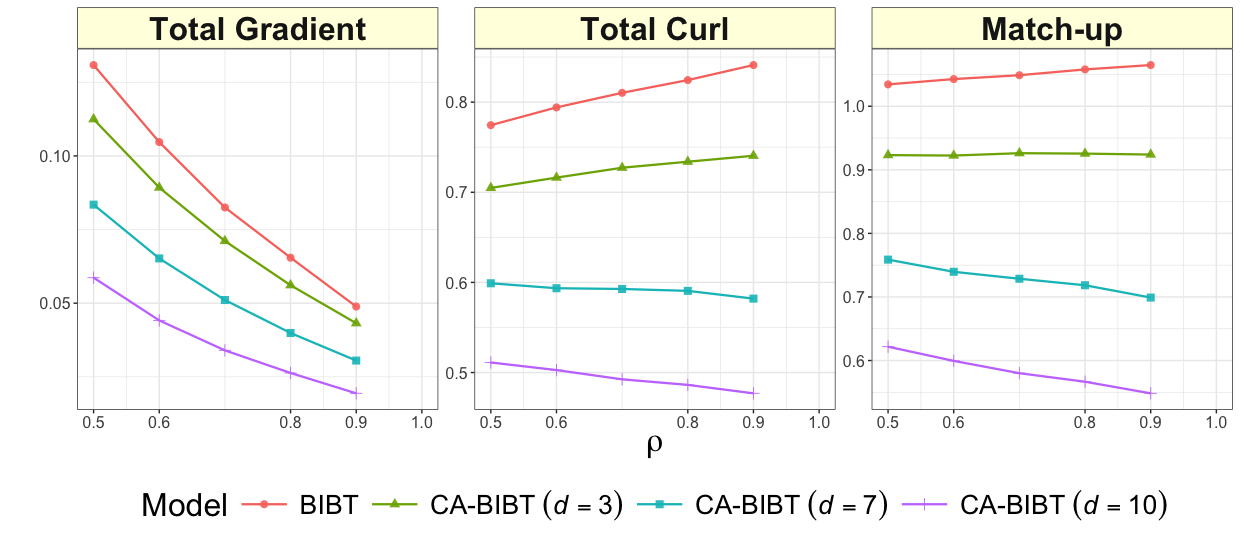}
        \caption{Evaluation on the missing edge set
        $\mathcal{E}_\mathrm{miss}$.}
        \label{fig:MSEs_incom_missing}
    \end{subfigure}

    \caption{
    Signal-normalized MSE as a function of $\rho$ for $N=20$ and $d_\mathrm{true}=10$.
    Panels~(a) and~(b) evaluate performance on the complete reference edge set $\mathcal{E}_\mathrm{com}$ and the missing edge set $\mathcal{E}_\mathrm{miss}$, respectively.
    Within each panel, the three component-specific plots correspond, from left to right, to the total gradient flow $\boldsymbol{M}_g$, the total curl flow $\boldsymbol{M}_c$, and the match-up vector $\boldsymbol{M}$.
    When $\rho=1$, $\mathcal{E}_\mathrm{miss}$ is empty, so the missing-edge metric is undefined and omitted.}
    \label{fig:MSEs_missing}
\end{figure}

\begin{figure}[htbp]
    \centering
    \begin{subfigure}[t]{0.8\linewidth}
        \centering
        \includegraphics[width=\linewidth]
        {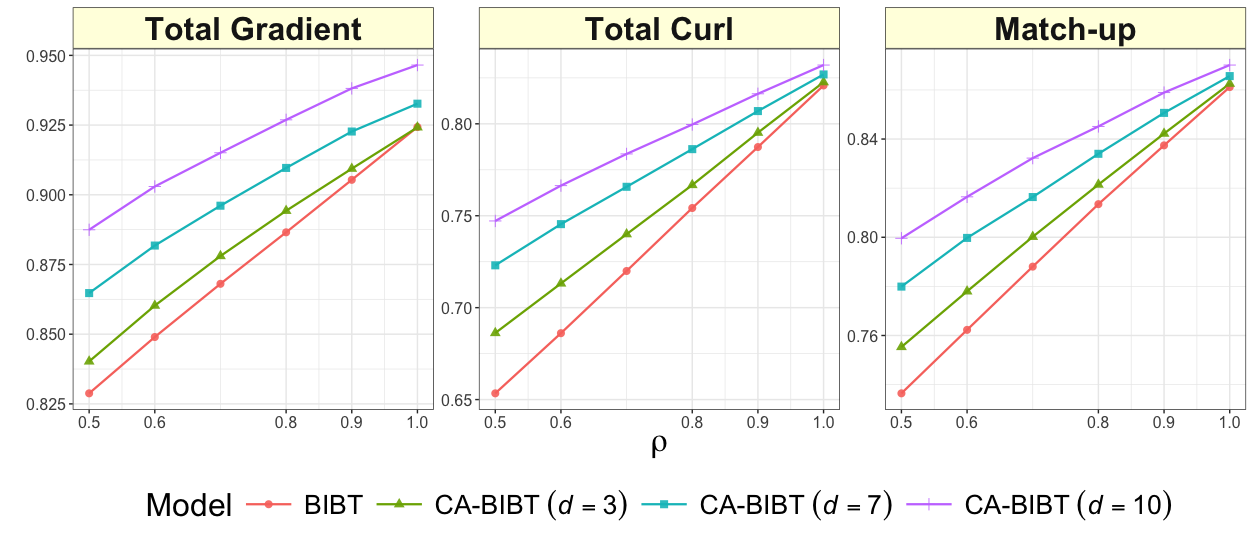}
        \caption{Evaluation on the complete reference edge set
        $\mathcal{E}_\mathrm{com}$.}
        \label{fig:Accuracies_incom_complete}
    \end{subfigure}

    \vspace{0.5em}

    \begin{subfigure}[t]{0.8\linewidth}
        \centering
        \includegraphics[width=\linewidth]
        {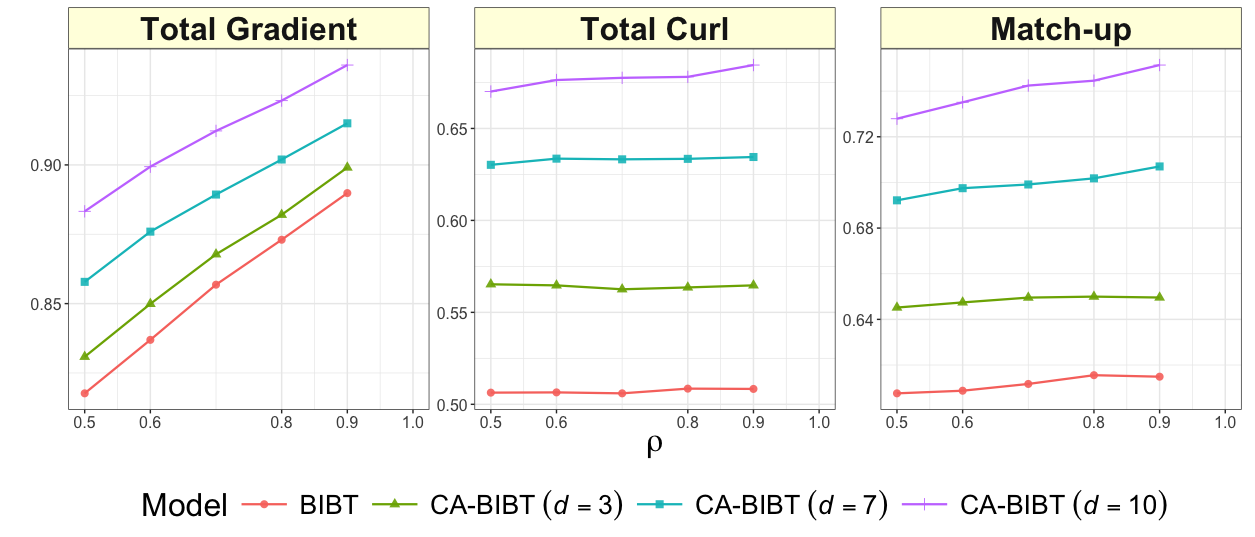}
        \caption{Evaluation on the missing edge set
        $\mathcal{E}_\mathrm{miss}$.}
        \label{fig:Accuracies_incom_missing}
    \end{subfigure}

    \caption{
    Sign-recovery accuracy as a function of $\rho$ for $N=20$ and $d_\mathrm{true}=10$.
    Panels~(a) and~(b) evaluate performance on the complete reference edge set $\mathcal{E}_\mathrm{com}$ and the missing edge set $\mathcal{E}_\mathrm{miss}$, respectively.
    Within each panel, the three component-specific plots correspond, from left to right, to the total gradient flow $\boldsymbol{M}_g$, the total curl flow $\boldsymbol{M}_c$, and the match-up vector $\boldsymbol{M}$.
    When $\rho=1$, $\mathcal{E}_\mathrm{miss}$ is empty, so the missing-edge metric is undefined and omitted.
    }
    \label{fig:Accuracies_missing}
\end{figure}

Figures~\ref{fig:MSEs_missing} and~\ref{fig:Accuracies_missing} summarize the effect of graph incompleteness on estimation error and sign recovery, respectively.
On the complete edge set $\mathcal{E}_\mathrm{com}$, performance deteriorates as the observation rate $\rho$ decreases, but CA-BIBT benefits from covariate information, reflecting its ability to explain part of the latent match-up structure through the covariate flow.
The behavior on the missing edge set $\mathcal{E}_\mathrm{miss}$ is component-specific.

The gradient component improves in both measures as $\rho$ increases to $1$ because missing gradient values are tied together through the shared vertex scores and can be extrapolated from the observed graph, provided that the observed graph remains connected.
By contrast, the performance of the total curl component is nearly insensitive to changes in $\rho$; this insensitivity is attributable to the residual curl component.
The residual curl flow can have degrees of freedom supported within $\mathcal{E}_\mathrm{miss}$ that vary independently of the observed edge set and are therefore unidentifiable from the observed likelihood.
Consequently, posterior behavior for the residual curl flow on missing edges is largely prior-dominated, and increasing $\rho$ does not necessarily improve either measure.

The covariate flow has a different extrapolation mechanism.
Although it is an edge flow, it is governed by the common low-dimensional coefficient vector $\boldsymbol{\beta}\in\mathbb{R}^d$.
If the covariate design restricted to the observed edges has full column rank, then the performance of the covariate flow improves as the observed comparison information grows, and the corresponding covariate flow on $\mathcal{E}_\mathrm{miss}$ is determined by the same coefficients with no additional structural degrees of freedom.
In the present simulation, the edge-specific covariates are generated from an independent Gaussian distribution, and the number of observed edges $|\mathcal{E}|\in\{95,\ldots,190\}$ is large relative to $d\in\{0,3,7,10\}$, so this rank condition is expected to hold with high probability.
This allows CA-BIBT to extrapolate covariate-induced gradient and curl structure to $\mathcal{E}_\mathrm{miss}$.
In empirical applications, however, this extrapolation may be weaker when some covariate directions are poorly represented on, or absent from, the observed graph.

Tables~\ref{tab:CP_CIL_Complete} and~\ref{tab:CP_CIL_Missing} complement the point-estimation results by reporting coverage probabilities and credible interval lengths for the complete and missing edge sets, $\mathcal{E}_\mathrm{com}$ and $\mathcal{E}_\mathrm{miss}$, respectively.
On $\mathcal{E}_\mathrm{com}$, the coverage probability of the total gradient flow generally improves as $\rho$ increases, whereas those of the total curl flow and the match-up vector remain comparatively stable; the credible interval lengths decrease for all three components.
On $\mathcal{E}_\mathrm{miss}$, the credible intervals also become shorter, but the coverage probabilities of the total curl flow and the match-up vector tend to decrease.
Because $|\mathcal{E}_\mathrm{miss}|$ decreases as $\rho$ increases, the coverage estimates on the missing edge set are based on fewer edges and hence become more variable at larger $\rho$.
Across all values of $\rho$, CA-BIBT yields consistently shorter credible intervals than BIBT for both edge sets, with the advantage becoming more pronounced as the covariate dimension increases.
These results indicate that the covariate-assisted structure increases posterior precision even for unobserved edges, as reflected in consistently shorter credible intervals.

\begin{table}[htbp]
    \centering
    \scriptsize
    \caption{Average coverage probabilities and credible interval lengths of $95\%$ posterior credible intervals for the complete reference edge set $\mathcal{E}_\mathrm{com}$ across different values of $\rho$.}
    \label{tab:CP_CIL_Complete}
    \resizebox{\textwidth}{!}{
    \begin{tabular}{ccccccccc}
        \toprule
        \multirow{3}{*}{Model}
        & \multirow{3}{*}{$\rho$}
        & \multicolumn{3}{c}{Coverage Probability}
        & \multicolumn{3}{c}{Credible Interval Length}
        & \multirow{3}{*}{\shortstack{Time\\(Seconds)}} \\
        \cmidrule(lr){3-5}
        \cmidrule(lr){6-8}
        & & \multirow{2}{*}{Match-up} & Total & Total
        & \multirow{2}{*}{Match-up} & Total & Total & \\
        & & & Gradient & Curl & & Gradient & Curl & \\
        \midrule
        \multirow{6}{*}{\shortstack{\textbf{BIBT}\\$(d=0)$}}
            & 0.5 & 0.920 & 0.933 & 0.911 & 2.61 & 1.26 & 2.37 & 8.0 \\
            & 0.6 & 0.919 & 0.929 & 0.910 & 2.39 & 1.10 & 2.20 & 8.2 \\
            & 0.7 & 0.918 & 0.928 & 0.910 & 2.18 & 0.95 & 2.03 & 8.5 \\
            & 0.8 & 0.918 & 0.927 & 0.911 & 1.99 & 0.82 & 1.86 & 8.7 \\
            & 0.9 & 0.920 & 0.936 & 0.913 & 1.81 & 0.71 & 1.69 & 9.1 \\
            & 1.0 & 0.920 & 0.948 & 0.915 & 1.65 & 0.60 & 1.53 & 9.2 \\
        \midrule
        \multirow{6}{*}{\shortstack{\textbf{CA-BIBT}\\$(d=3)$}}
            & 0.5 & 0.919 & 0.928 & 0.909 & 2.47 & 1.15 & 2.25 & 8.0 \\
            & 0.6 & 0.916 & 0.922 & 0.908 & 2.26 & 0.99 & 2.09 & 8.2 \\
            & 0.7 & 0.915 & 0.917 & 0.907 & 2.07 & 0.86 & 1.93 & 8.4 \\
            & 0.8 & 0.915 & 0.917 & 0.910 & 1.90 & 0.75 & 1.78 & 8.6 \\
            & 0.9 & 0.916 & 0.922 & 0.911 & 1.75 & 0.65 & 1.64 & 8.8 \\
            & 1.0 & 0.915 & 0.930 & 0.912 & 1.60 & 0.56 & 1.50 & 9.0 \\
        \midrule
        \multirow{6}{*}{\shortstack{\textbf{CA-BIBT}\\$(d=7)$}}
            & 0.5 & 0.915 & 0.923 & 0.908 & 2.28 & 0.98 & 2.10 & 7.8 \\
            & 0.6 & 0.914 & 0.919 & 0.908 & 2.09 & 0.84 & 1.95 & 8.0 \\
            & 0.7 & 0.912 & 0.916 & 0.906 & 1.92 & 0.73 & 1.81 & 8.2 \\
            & 0.8 & 0.911 & 0.916 & 0.907 & 1.78 & 0.64 & 1.67 & 8.5 \\
            & 0.9 & 0.912 & 0.921 & 0.908 & 1.65 & 0.56 & 1.56 & 8.7 \\
            & 1.0 & 0.911 & 0.924 & 0.908 & 1.53 & 0.50 & 1.44 & 9.0 \\
        \midrule
        \multirow{6}{*}{\shortstack{\textbf{CA-BIBT}\\$(d=10)$}}
            & 0.5 & 0.915 & 0.929 & 0.908 & 2.12 & 0.84 & 1.98 & 7.7 \\
            & 0.6 & 0.913 & 0.929 & 0.907 & 1.95 & 0.73 & 1.83 & 8.0 \\
            & 0.7 & 0.911 & 0.929 & 0.905 & 1.80 & 0.63 & 1.70 & 8.3 \\
            & 0.8 & 0.909 & 0.933 & 0.904 & 1.67 & 0.56 & 1.59 & 8.6 \\
            & 0.9 & 0.910 & 0.942 & 0.906 & 1.56 & 0.49 & 1.49 & 8.9 \\
            & 1.0 & 0.910 & 0.949 & 0.906 & 1.46 & 0.44 & 1.39 & 9.0 \\
        \bottomrule
    \end{tabular}
    }
\end{table}

\begin{table}[htbp]
    \centering
    \scriptsize
    \caption{Average coverage probabilities and credible interval lengths of $95\%$ posterior credible intervals for the missing edge set $\mathcal{E}_\mathrm{miss}$ across different values of $\rho$.
    Results for $\rho=1$ are omitted because $\mathcal{E}_\mathrm{miss}$ is empty.}    
    \label{tab:CP_CIL_Missing}
    \resizebox{\textwidth}{!}{
    \begin{tabular}{cccccccc}
        \toprule
        \multirow{3}{*}{Model} & \multirow{3}{*}{$\rho$} & \multicolumn{3}{c}{Coverage Probability} & \multicolumn{3}{c}{Credible Interval Length} \\
        \cmidrule(lr){3-5} \cmidrule(lr){6-8}
        & & \multirow{2}{*}{Match-up} & Total & Total & \multirow{2}{*}{Match-up} & Total & Total \\
        & & & Gradient & Curl & & Gradient & Curl \\
        \midrule
        \multirow{5}{*}{\shortstack{\textbf{BIBT}\\$(d=0)$}}
            & 0.5 & 0.903 & 0.934 & 0.897 & 3.51 & 1.34 & 2.99 \\
            & 0.6 & 0.894 & 0.929 & 0.890 & 3.43 & 1.18 & 2.96 \\
            & 0.7 & 0.886 & 0.927 & 0.883 & 3.35 & 1.03 & 2.92 \\
            & 0.8 & 0.877 & 0.923 & 0.875 & 3.27 & 0.91 & 2.87 \\
            & 0.9 & 0.868 & 0.928 & 0.866 & 3.20 & 0.80 & 2.82 \\
        \midrule
        \multirow{5}{*}{\shortstack{\textbf{CA-BIBT}\\$(d=3)$}}
            & 0.5 & 0.901 & 0.927 & 0.894 & 3.26 & 1.21 & 2.81 \\
            & 0.6 & 0.892 & 0.920 & 0.887 & 3.16 & 1.05 & 2.76 \\
            & 0.7 & 0.881 & 0.915 & 0.878 & 3.06 & 0.92 & 2.70 \\
            & 0.8 & 0.872 & 0.912 & 0.870 & 2.98 & 0.81 & 2.65 \\
            & 0.9 & 0.863 & 0.914 & 0.863 & 2.91 & 0.71 & 2.60 \\
        \midrule
        \multirow{5}{*}{\shortstack{\textbf{CA-BIBT}\\$(d=7)$}}
            & 0.5 & 0.899 & 0.922 & 0.894 & 2.93 & 1.02 & 2.59 \\
            & 0.6 & 0.891 & 0.917 & 0.889 & 2.83 & 0.88 & 2.52 \\
            & 0.7 & 0.883 & 0.912 & 0.879 & 2.72 & 0.77 & 2.45 \\
            & 0.8 & 0.874 & 0.910 & 0.872 & 2.63 & 0.68 & 2.39 \\
            & 0.9 & 0.871 & 0.912 & 0.868 & 2.56 & 0.60 & 2.33 \\
        \midrule
        \multirow{5}{*}{\shortstack{\textbf{CA-BIBT}\\$(d=10)$}}
            & 0.5 & 0.900 & 0.929 & 0.896 & 2.67 & 0.87 & 2.40 \\
            & 0.6 & 0.892 & 0.928 & 0.888 & 2.56 & 0.75 & 2.32 \\
            & 0.7 & 0.883 & 0.928 & 0.881 & 2.46 & 0.66 & 2.25 \\
            & 0.8 & 0.875 & 0.931 & 0.870 & 2.36 & 0.58 & 2.17 \\
            & 0.9 & 0.869 & 0.939 & 0.866 & 2.29 & 0.52 & 2.12 \\
        \bottomrule
    \end{tabular}
    }
\end{table}

\subsection{Sensitivity Analysis for the Residual Curl Shrinkage Prior}
\label{sup:sensitivity}
Because the dimension $q_z$ of the residual curl coefficient vector $\boldsymbol{z}\in\mathbb{R}^{q_z}$ can be large, we assess the sensitivity of the posterior conclusions to the shrinkage prior placed on its components.
We revisit both empirical analyses of Section~6: the canary dominance analysis under CA-BIBT and the guanaco dominance analysis under the covariate-free BIBT.
The aim of this analysis is not to compare the operating characteristics of competing shrinkage priors, but to assess whether the substantive conclusions of Sections~6.1 and~6.2 are robust to the choice of prior for the residual curl component.

For $\ell=1,\ldots,q_z$, the horseshoe prior \citep{carvalho2010Horseshoe} placed on $z_\ell$ is
\begin{gather}
    z_\ell \mid \tau,\lambda_\ell \sim N(0,\tau^2\lambda_\ell^2),\\
    \lambda_\ell \sim C^+(0,1),\quad \tau \sim C^+(0,1),
\end{gather}
where $\lambda_\ell$ and $\tau$ are the local and global shrinkage parameters, respectively.
The local-scale prior $\lambda_\ell\sim C^+(0,1)$ is equivalent to $\lambda_\ell^2 \sim \mathrm{BetaPrime}(1/2, 1/2)$.
Following the generalized beta mixture of Gaussians \citep{armagan2011Generalized}, we embed this specification in the wider family $\lambda_\ell^2\sim \mathrm{BetaPrime}(a,b)$ with $a,b>0$.
Smaller values of $a$ place more prior mass near $\lambda_\ell^2=0$ and thus induce stronger shrinkage toward zero, whereas smaller values of $b$ yield heavier tails and allow strongly supported residual curl directions to escape shrinkage.

For posterior computation, we adopt an inverse gamma mixture of inverse gamma distributions \citep{makalic2016Simple, schmidt2020Bayesian},
\begin{gather}
    z_\ell \mid \tau, \lambda_\ell \sim N(0, \tau^2 \lambda_\ell^2),\\
    \lambda_\ell^2 \mid \nu_\ell \sim \mathrm{IG}\!\left(b, \tfrac{1}{\nu_\ell}\right),\qquad
    \nu_\ell \sim \mathrm{IG}(a,1),\\
    \tau^2 \mid \xi \sim \mathrm{IG}\!\left(\tfrac{1}{2}, \tfrac{1}{\xi}\right),\qquad
    \xi \sim \mathrm{IG}\!\left(\tfrac{1}{2},1\right),
\end{gather}
which implies $\lambda_\ell^2\sim \mathrm{BetaPrime}(a,b)$ and $\tau\sim C^+(0,1)$, and recovers the standard horseshoe prior at $a=b=1/2$.

Because only the local-scale hierarchy is modified, the generalized prior leaves the Gibbs sampler of Section~3.3 unchanged except for the full conditionals of $\lambda_\ell^2$ and $\nu_\ell$.
Combining the Gaussian likelihood contribution of $z_\ell$ with the inverse-gamma priors above yields the conjugate updates:
\begin{itemize}
    \item[-] \textbf{(Update of $\lambda_\ell^2$)}\
    Generate $\lambda_\ell^2$ from $\mathrm{IG}(a_\lambda^{(\ell)}, b_\lambda^{(\ell)})$, where
    \begin{equation}
        a_\lambda^{(\ell)} = b + \frac{1}{2},\quad
        b_\lambda^{(\ell)} = \frac{1}{\nu_\ell} + \frac{z_\ell^2}{2\tau^2}.
    \end{equation}
    \item[-] \textbf{(Update of $\nu_\ell$)}\
    Generate $\nu_\ell$ from $\mathrm{IG}(a_\nu^{(\ell)}, b_\nu^{(\ell)})$, where
    $a_\nu^{(\ell)} = a+b$ and $b_\nu^{(\ell)} = 1 + 1/\lambda_\ell^2$.
\end{itemize}
The updates for $\tau^2$ and $\xi$ are identical to those in Section~3.3.
Setting $a=b=1/2$ reduces the shape parameters to $b+1/2=1$ and $a+b=1$, recovering the standard horseshoe updates and confirming consistency with the baseline sampler.
Consequently, the sensitivity analysis incurs no additional computational cost relative to the main specification.

We vary $a,b\in\{1/4,1/2,3/4\}$, giving nine configurations that span stronger-to-weaker shrinkage (smaller-to-larger $a$) and heavier-to-lighter tails (smaller-to-larger $b$), with the standard horseshoe at $a=b=1/2$.
For each configuration, we refit each model under the same MCMC settings as in Section~6 and examine the key posterior summaries, namely the flow contribution ratios, the posterior probabilities of the stochastic transitivity (ST) classes, the BFDR-calibrated decision summaries, and the covariate coefficients for CA-BIBT.

Tables~\ref{tab:sensitivity_canary} and~\ref{tab:sensitivity_guanaco} show that the qualitative conclusions of both analyses are preserved across all nine configurations.
For the canary data, the flow contribution ratios vary only slightly across the nine configurations, the covariate contribution remains concentrated in the gradient subspace with a moderate contribution to the curl flow space, and the posterior probability of SIT remains $\pi_\mathrm{I}=1$ throughout.
The covariate coefficients show equally stable behavior in Table~\ref{tab:sensitivity_canary_beta}.
The male-dominance interpretation of Section~6.1 is therefore unaffected by the choice of shrinkage prior.

\begin{table}[htbp]
\centering
\scriptsize
\caption{Posterior means of the four flow contribution ratios for the canary dominance data under nine local-scale prior configurations $\lambda_\ell^2\sim\mathrm{BetaPrime}(a,b)$.}
\label{tab:sensitivity_canary}
\resizebox{\textwidth}{!}{
\begin{tabular}{c ccc ccc ccc}
\toprule 
 & \multicolumn{3}{c}{$a{=}1/4$}
 & \multicolumn{3}{c}{$a{=}1/2$}
 & \multicolumn{3}{c}{$a{=}3/4$} \\
 \cmidrule(lr){2-4}\cmidrule(lr){5-7}\cmidrule(lr){8-10}
 Summary
 & $b{=}1/4$ & $b{=}1/2$ & $b{=}3/4$
 & $b{=}1/4$ & $b{=}1/2$ & $b{=}3/4$
 & $b{=}1/4$ & $b{=}1/2$ & $b{=}3/4$ \\
\midrule
$R_g$         & $0.568$   & $0.567$   & $0.568$   & $0.568$   & $0.568$   & $0.567$   & $0.568$   & $0.568$   & $0.567$ \\
$R_c$         & $0.432$   & $0.433$   & $0.432$   & $0.432$   & $0.432$   & $0.433$   & $0.432$   & $0.432$   & $0.433$ \\
$R_{x\mid g}$ & $0.822$   & $0.821$   & $0.822$   & $0.820$   & $0.819$   & $0.818$   & $0.819$   & $0.817$   & $0.817$ \\
$R_{x\mid c}$ & $0.340$   & $0.339$   & $0.340$   & $0.339$   & $0.340$   & $0.338$   & $0.340$   & $0.339$   & $0.339$ \\
\bottomrule
\end{tabular}
}
\end{table}

\begin{table}[htbp]
\centering
\scriptsize
\caption{Posterior summaries of the covariate coefficients for the canary
dominance data under the nine local-scale prior configurations
$\lambda_\ell^2\sim\mathrm{BetaPrime}(a,b)$, where $\beta_1$ and $\beta_2$ denote
the sex and mate effects defined in Section~6.1. Credible intervals are
equal-tailed $95\%$ intervals and exclude zero in every configuration.}
\label{tab:sensitivity_canary_beta}
\resizebox{\textwidth}{!}{
\begin{tabular}{cc cccc cccc}
\toprule
 & & \multicolumn{4}{c}{$\beta_1$ (Sex Effect)} & \multicolumn{4}{c}{$\beta_2$ (Mate Effect)} \\
\cmidrule(lr){3-6}\cmidrule(lr){7-10}
$a$ & $b$ & Mean & Median & SD & $95\%$ CI & Mean & Median & SD & $95\%$ CI \\
\midrule
$1/4$ & $1/4$ & $1.779$ & $1.778$ & $0.076$ & $[1.635,\,1.931]$ & $-1.948$ & $-1.948$ & $0.112$ & $[-2.171,\,-1.731]$ \\
$1/4$ & $1/2$ & $1.772$ & $1.768$ & $0.075$ & $[1.628,\,1.922]$ & $-1.939$ & $-1.940$ & $0.112$ & $[-2.158,\,-1.723]$ \\
$1/4$ & $3/4$ & $1.773$ & $1.772$ & $0.076$ & $[1.630,\,1.920]$ & $-1.941$ & $-1.940$ & $0.112$ & $[-2.159,\,-1.720]$ \\
$1/2$ & $1/4$ & $1.775$ & $1.774$ & $0.076$ & $[1.629,\,1.927]$ & $-1.942$ & $-1.943$ & $0.112$ & $[-2.176,\,-1.735]$ \\
$1/2$ & $1/2$ & $1.769$ & $1.769$ & $0.077$ & $[1.623,\,1.925]$ & $-1.939$ & $-1.938$ & $0.115$ & $[-2.167,\,-1.719]$ \\
$1/2$ & $3/4$ & $1.769$ & $1.768$ & $0.078$ & $[1.622,\,1.924]$ & $-1.937$ & $-1.936$ & $0.114$ & $[-2.173,\,-1.731]$ \\
$3/4$ & $1/4$ & $1.774$ & $1.774$ & $0.075$ & $[1.626,\,1.919]$ & $-1.943$ & $-1.943$ & $0.112$ & $[-2.158,\,-1.721]$ \\
$3/4$ & $1/2$ & $1.771$ & $1.771$ & $0.074$ & $[1.626,\,1.914]$ & $-1.940$ & $-1.940$ & $0.111$ & $[-2.144,\,-1.714]$ \\
$3/4$ & $3/4$ & $1.765$ & $1.764$ & $0.075$ & $[1.625,\,1.917]$ & $-1.936$ & $-1.935$ & $0.112$ & $[-2.160,\,-1.723]$ \\
\bottomrule
\end{tabular}
}
\end{table}

For the guanaco data, the principal conclusions are robust to the shrinkage prior, although the posterior summaries show some quantitative variation.
Across all prior specifications, the gradient-dominant structure and the conclusion that WST is plausible but not decisively supported remain unchanged.
The high-credibility features of the BFDR-calibrated decision summaries are also stable.
The only exception is that N.8 precedes Pink under $(a,b)=(3/4,1/2)$; however, the posterior match-up for this pair is centered near zero, with a directional probability close to $1/2$ under both this prior and the standard horseshoe prior.
This reversal therefore reflects intrinsic uncertainty about this particular edge rather than a substantive change in the dominance structure.
Accordingly, the coarser rankings obtained at larger $\epsilon$, or under more stringent BFDR calibration, are robust to the prior, whereas finer distinctions near $\epsilon=1/2$ remain prior-sensitive.
We do not interpret these results as evidence that the horseshoe prior is optimal, but rather as evidence that adopting this well-established shrinkage prior for sparse high-dimensional coefficients \citep{carvalho2009Handling, carvalho2010Horseshoe} does not materially affect the conclusions drawn here.

The larger quantitative sensitivity of the guanaco analysis relative to the canary analysis is consistent with greater prior exposure when the likelihood is weakly informative about the latent match-up values, which may arise from limited pairwise replication, incomplete observation designs, or weak and localized latent signal structure.
Because the two datasets differ in several respects beyond the number of observed comparisons $n_{ij}$, this comparison should be read as descriptive rather than as a controlled characterization of prior sensitivity.
A systematic assessment of how prior sensitivity depends jointly on the comparison design and the latent signal structure is left to future controlled simulation studies.

\begin{table}[htbp]
\centering
\scriptsize
\caption{Posterior means of the gradient and curl contribution ratios and the posterior probabilities of the stochastic transitivity classes for the guanaco dominance data under nine local-scale prior configurations $\lambda_\ell^2\sim\mathrm{BetaPrime}(a,b)$.
The guanaco analysis uses the covariate-free BIBT, so the within-subspace covariate contribution ratios are not defined.}
\label{tab:sensitivity_guanaco}
\resizebox{\textwidth}{!}{
\begin{tabular}{c ccc ccc ccc}
\toprule
 & \multicolumn{3}{c}{$a{=}1/4$}
 & \multicolumn{3}{c}{$a{=}1/2$}
 & \multicolumn{3}{c}{$a{=}3/4$} \\
\cmidrule(lr){2-4}\cmidrule(lr){5-7}\cmidrule(lr){8-10}
Summary
 & $b{=}1/4$ & $b{=}1/2$ & $b{=}3/4$
 & $b{=}1/4$ & $b{=}1/2$ & $b{=}3/4$
 & $b{=}1/4$ & $b{=}1/2$ & $b{=}3/4$ \\
\midrule
$R_g$   & $0.938$   & $0.948$   & $0.942$   & $0.956$   & $0.949$   & $0.948$   & $0.939$   & $0.938$   & $0.944$ \\
$R_c$   & $0.063$   & $0.052$   & $0.058$   & $0.044$   & $0.051$   & $0.052$   & $0.061$   & $0.062$   & $0.056$ \\
$\pi_\mathrm{S}$ & $0.012$   & $0.070$   & $0.026$   & $0.068$   & $0.043$   & $0.028$   & $0.024$   & $0.032$   & $0.030$ \\
$\pi_\mathrm{M}$ & $0.247$   & $0.366$   & $0.289$   & $0.402$   & $0.341$   & $0.318$   & $0.258$   & $0.285$   & $0.305$ \\
$\pi_\mathrm{W}$ & $0.557$   & $0.628$   & $0.589$   & $0.675$   & $0.627$   & $0.614$   & $0.570$   & $0.573$   & $0.592$ \\
$\pi_\mathrm{I}$ & $0.443$   & $0.372$   & $0.411$   & $0.325$   & $0.373$   & $0.386$   & $0.430$   & $0.427$   & $0.408$ \\
\bottomrule
\end{tabular}
}
\end{table}

\section{Proofs}
\label{supplementary:proofs}
\subsection*{Proof of Theorem~1}
Suppose that two parameter triples $(\boldsymbol{s},\boldsymbol{\Phi},\boldsymbol{\beta})$ and $(\boldsymbol{s}',\boldsymbol{\Phi}',\boldsymbol{\beta}')$ in $\Theta(\boldsymbol{v},V,X_E)$ yield the same edge probabilities. 
By the strict monotonicity of $\sigma$, we have
\begin{equation}
\label{sup:identifiability_equation}
    G(\boldsymbol{s}-\boldsymbol{s}') + C^\top(\boldsymbol{\Phi}-\boldsymbol{\Phi}') + X_E^\top(\boldsymbol{\beta}-\boldsymbol{\beta}') = \boldsymbol{0}.
\end{equation}

Let $P_\mathrm{grad}$ and $P_\mathrm{curl}$ denote the orthogonal projections in $L^2_\wedge(\mathcal{E})$ onto $\mathrm{im}(\mathrm{grad})$ and $\mathrm{im}(\mathrm{curl}^\ast)$, respectively.
Then,
\begin{equation}
    P_\mathrm{grad}=G(G^\top G)^+G^\top,\quad
    P_\mathrm{curl}=C^\top(CC^\top)^+C,
\end{equation}
where $+$ denotes the Moore--Penrose inverse.
By the orthogonal Hodge decomposition, $P_\mathrm{grad}+P_\mathrm{curl}=I$ and
\begin{equation}
    P_\mathrm{grad}G=G,\quad
    P_\mathrm{curl}G=0,\quad
    P_\mathrm{grad}C^\top=0,\quad
    P_\mathrm{curl}C^\top=C^\top.
\end{equation}

Applying $P_\mathrm{grad}$ and $P_\mathrm{curl}$ to \eqref{sup:identifiability_equation} gives
\begin{align}
\label{sup:grad_equation}
    G(\boldsymbol{s}-\boldsymbol{s}') + P_\mathrm{grad} X_E^\top(\boldsymbol{\beta}-\boldsymbol{\beta}') &= \boldsymbol{0}, \\
\label{sup:curl_equation}
    C^\top(\boldsymbol{\Phi}-\boldsymbol{\Phi}') + P_\mathrm{curl} X_E^\top(\boldsymbol{\beta}-\boldsymbol{\beta}') &= \boldsymbol{0}.
\end{align}

Taking the inner product of \eqref{sup:grad_equation} with $G(\boldsymbol{s}-\boldsymbol{s}')$ yields
\begin{equation}
    \|G(\boldsymbol{s}-\boldsymbol{s}')\|_2^2
    + \left\langle P_{\mathrm{grad}} X_E^\top(\boldsymbol{\beta}-\boldsymbol{\beta}'), G(\boldsymbol{s}-\boldsymbol{s}') \right\rangle_\mathcal{E}
    = 0.
\end{equation}
Since $P_{\mathrm{grad}}$ is self-adjoint and $G(\boldsymbol{s}-\boldsymbol{s}') \in \mathrm{im}(\mathrm{grad})$, we have
\begin{equation}
    \left\langle P_{\mathrm{grad}} X_E^\top(\boldsymbol{\beta}-\boldsymbol{\beta}'), G(\boldsymbol{s}-\boldsymbol{s}') \right\rangle_\mathcal{E}
    = (\boldsymbol{\beta}-\boldsymbol{\beta}')^\top
    X_E G(\boldsymbol{s}-\boldsymbol{s}').
\end{equation}
By the constraints $X_E G\boldsymbol{s} = X_E G\boldsymbol{s}' = \boldsymbol{0}$, it follows that $\|G(\boldsymbol{s}-\boldsymbol{s}')\|_2^2 = 0$, and hence $G(\boldsymbol{s}-\boldsymbol{s}') = \boldsymbol{0}$.
Substituting back into \eqref{sup:grad_equation} gives $P_\mathrm{grad} X_E^\top(\boldsymbol{\beta}-\boldsymbol{\beta}') = \boldsymbol{0}$.

Similarly, taking the inner product of \eqref{sup:curl_equation} with $C^\top(\boldsymbol{\Phi}-\boldsymbol{\Phi}')$ yields 
\begin{equation}
    \|C^\top (\boldsymbol{\Phi}-\boldsymbol{\Phi}')\|_2^2
    + \left\langle P_{\mathrm{curl}} X_E^\top(\boldsymbol{\beta}-\boldsymbol{\beta}'), C^\top (\boldsymbol{\Phi}-\boldsymbol{\Phi}') \right\rangle_\mathcal{E}
    = 0.
\end{equation}
Since $P_{\mathrm{curl}}$ is self-adjoint and $C^\top (\boldsymbol{\Phi}-\boldsymbol{\Phi}') \in \mathrm{im}(\mathrm{curl}^\ast)$, we have
\begin{equation}
    \left\langle P_{\mathrm{curl}} X_E^\top(\boldsymbol{\beta}-\boldsymbol{\beta}'), C^\top (\boldsymbol{\Phi}-\boldsymbol{\Phi}') \right\rangle_\mathcal{E}
    = (\boldsymbol{\beta}-\boldsymbol{\beta}')^\top
    X_E C^\top (\boldsymbol{\Phi}-\boldsymbol{\Phi}').
\end{equation}
By the constraints $X_E C^\top \boldsymbol{\Phi} = X_E C^\top \boldsymbol{\Phi}' = \boldsymbol{0}$, it follows that $\|C^\top (\boldsymbol{\Phi}-\boldsymbol{\Phi}')\|_2^2 = 0$, and hence $C^\top (\boldsymbol{\Phi}-\boldsymbol{\Phi}') = \boldsymbol{0}$.
Substituting back into \eqref{sup:curl_equation} gives $P_\mathrm{curl} X_E^\top(\boldsymbol{\beta}-\boldsymbol{\beta}') = \boldsymbol{0}$.

Summing the two resulting equations and using $P_\mathrm{grad} + P_\mathrm{curl} = I$ gives $X_E^\top(\boldsymbol{\beta}-\boldsymbol{\beta}') = \boldsymbol{0}$, and since $X_E$ has full row rank, this implies $\boldsymbol{\beta} = \boldsymbol{\beta}'$.

It remains to show $\boldsymbol{s}=\boldsymbol{s}'$ and $\boldsymbol{\Phi}=\boldsymbol{\Phi}'$.
From the preceding argument, we already have
\begin{equation}
    G(\boldsymbol{s}-\boldsymbol{s}')=\boldsymbol{0},\quad
    C^\top(\boldsymbol{\Phi}-\boldsymbol{\Phi}')=\boldsymbol{0}.
\end{equation}
Since $\ker(\mathrm{grad}) = \mathrm{span}(\boldsymbol{1})$, there exists $c \in \mathbb{R}$ such that $\boldsymbol{s} - \boldsymbol{s}' = c \boldsymbol{1}$.
Using the constraints $\boldsymbol{v}^\top\boldsymbol{s}=\boldsymbol{v}^\top\boldsymbol{s}'=0$, we obtain
\begin{equation}
    0 = \boldsymbol{v}^\top \boldsymbol{s} - \boldsymbol{v}^\top \boldsymbol{s}' 
      = \boldsymbol{v}^\top (\boldsymbol{s} - \boldsymbol{s}') 
      = c \boldsymbol{v}^\top \boldsymbol{1}.
\end{equation}
By assumption, $\boldsymbol{v}^\top \boldsymbol{1} \neq 0$, and hence $c=0$. 
Therefore, $\boldsymbol{s} = \boldsymbol{s}'$.

Similarly, the second equality implies $\boldsymbol{\Phi}-\boldsymbol{\Phi}'\in\ker(\mathrm{curl}^\ast)$.
Since the columns of $A$ form a basis of $\ker(\mathrm{curl}^\ast)$, there exists $\boldsymbol{b}\in\mathbb{R}^{|\mathcal{T}|-K}$ such that $\boldsymbol{\Phi}-\boldsymbol{\Phi}'=A\boldsymbol{b}$.
Using the constraints $V^\top\boldsymbol{\Phi}=V^\top\boldsymbol{\Phi}'=\boldsymbol{0}$, we obtain
\begin{equation}
    0 = V^\top \boldsymbol{\Phi} - V^\top \boldsymbol{\Phi}'
      = V^\top (\boldsymbol{\Phi} - \boldsymbol{\Phi}') 
      = V^\top A \boldsymbol{b}.
\end{equation}
Since $\det(V^\top A)\neq0$, the matrix $V^\top A$ is nonsingular, and hence $\boldsymbol{b}=\boldsymbol{0}$. Therefore, $\boldsymbol{\Phi}=\boldsymbol{\Phi}'$.

Together with $\boldsymbol{\beta}=\boldsymbol{\beta}'$, we conclude that
\begin{equation}
    (\boldsymbol{s},\boldsymbol{\Phi},\boldsymbol{\beta}) = (\boldsymbol{s}',\boldsymbol{\Phi}',\boldsymbol{\beta}').
\end{equation}
Therefore, the mapping
\begin{equation}
    (\boldsymbol{s},\boldsymbol{\Phi},\boldsymbol{\beta})
    \in\Theta(\boldsymbol{v},V,X_E)
    \mapsto
    \{\sigma(M_{ij})\}_{\{i,j\}\in\mathcal{E}}
\end{equation}
is injective, and the CA-BIBT model is identifiable over $\Theta(\boldsymbol{v},V,X_E)$. \qed

\subsection*{Proof of Proposition~1}
\textbf{Existence of a finest blockwise ranking}:
The trivial partition $(\mathcal{V})$ is always a blockwise ranking, and the number of blocks in any ordered partition of $\mathcal{V}$ is an integer between $1$ and $|\mathcal{V}|$. 
Hence, a blockwise ranking with the largest possible number of blocks exists.

\textbf{Uniqueness of the finest blockwise ranking}:
Let $\mathcal{Q}=(Q_1,\ldots,Q_g)$ and $\mathcal{W}=(W_1,\ldots,W_h)$ be two blockwise rankings with respect to $R$. 
Consider the set of all nonempty intersections 
\begin{equation}
    \mathcal{S} = \{Q_s\cap W_t \mid Q_s\cap W_t\neq\emptyset\}.
\end{equation}
We claim that these intersections can be ordered consistently.
Suppose that $Q_s \cap W_t \neq\emptyset$ and $Q_{s'}\cap W_{t'}\neq\emptyset$ with $s<s'$ and $t>t'$. 
For any $i\in Q_s\cap W_t$ and $j\in Q_{s'}\cap W_{t'}$, the blockwise property of $\mathcal{Q}$ gives $i\mathrel{R}j$, whereas that of $\mathcal{W}$ gives $j\mathrel{R}i$, contradicting the asymmetry of $R$.
Hence, no such crossing can occur.

Therefore, the nonempty intersections may be ordered lexicographically by $(s,t)$.
Let $T_{st}=Q_s\cap W_t$ and $T_{s't'}=Q_{s'}\cap W_{t'}$ be two such blocks with $(s,t)$ preceding $(s',t')$.
If $s<s'$, then $i\mathrel{R}j$ for all $i\in T_{st}$ and $j\in T_{s't'}$, by the blockwise property of $\mathcal{Q}$.
If $s=s'$, then necessarily $t<t'$, and the same conclusion follows from the blockwise property of $\mathcal{W}$.
Thus, $\mathcal{S}$ is itself a blockwise ranking.

Now suppose that $\mathcal{Q}$ and $\mathcal{W}$ both have the largest possible number of blocks.
If their underlying collections of blocks are different, then at least one block of $\mathcal{Q}$ is split by $\mathcal{W}$, or vice versa.
Hence, their common refinement $\mathcal{S}$ has strictly more blocks than $\mathcal{Q}$, which contradicts the assumption that $\mathcal{Q}$ attains the largest possible number of blocks.
Thus, $\mathcal{Q}$ and $\mathcal{W}$ have the same blocks.
Their order must also coincide: if two blocks were ordered differently in $\mathcal{Q}$ and $\mathcal{W}$, the blockwise property would imply both $i\mathrel{R}j$ and $j\mathrel{R}i$ for some $i,j$, contradicting the asymmetry of $R$.
Therefore, $\mathcal{Q} = \mathcal{W}$. \qed

\subsection*{Proof of Theorem~\ref{thm:identifiability_BIBT}}
Suppose $(\boldsymbol{s},\boldsymbol{\Phi}), (\boldsymbol{s}',\boldsymbol{\Phi}') \in\Theta(\boldsymbol{v},V)$ induce the same edge probabilities.
By the strict monotonicity of $\sigma$, $G(\boldsymbol{s}-\boldsymbol{s}') + C^\top(\boldsymbol{\Phi}-\boldsymbol{\Phi}')=\boldsymbol{0}$.
Since $\mathrm{im}(\mathrm{grad})$ and $\mathrm{im}(\mathrm{curl}^\ast)$ are orthogonal in $L^2_\wedge(\mathcal{E})$, the two summands vanish separately: $G(\boldsymbol{s}-\boldsymbol{s}')=\boldsymbol{0}$ and $C^\top(\boldsymbol{\Phi}-\boldsymbol{\Phi}')=\boldsymbol{0}$.
The first gives $\boldsymbol{s}-\boldsymbol{s}'\in\ker(\mathrm{grad})=\mathrm{span}(\boldsymbol{1})$, so $\boldsymbol{s}-\boldsymbol{s}'=c\boldsymbol{1}$; the constraint $\boldsymbol{v}^\top(\boldsymbol{s}-\boldsymbol{s}')=c\,\boldsymbol{v}^\top\boldsymbol{1}=0$ with $\boldsymbol{v}^\top\boldsymbol{1}\neq0$ forces $c=0$, and hence $\boldsymbol{s}=\boldsymbol{s}'$.
The second gives $\boldsymbol{\Phi}-\boldsymbol{\Phi}'=A\boldsymbol{b}$ for some $\boldsymbol{b}$; the constraint $V^\top(\boldsymbol{\Phi}-\boldsymbol{\Phi}')=(V^\top A)\boldsymbol{b}=\boldsymbol{0}$ with $\det(V^\top A)\neq0$ forces $\boldsymbol{b}=\boldsymbol{0}$, and hence $\boldsymbol{\Phi}=\boldsymbol{\Phi}'$.
Therefore, we establish the injectivity of $(\boldsymbol{s},\boldsymbol{\Phi})\mapsto \{\sigma(M_{ij})\}_{\{i,j\}\in\mathcal{E}}$ over $\Theta(\boldsymbol{v},V)$. \qed

\subsection*{Proof of Proposition~\ref{prp:BFDR-blockwise}}
\textbf{Nonemptiness of $\mathfrak{B}_\alpha$}:
If $\max_{i\neq j}q_{ij}<1$, then choosing $\epsilon\in [\max_{i\neq j} q_{ij}, 1)$ makes the relation $\to_\epsilon$ empty.
Hence, $\mathcal{B}(\to_\epsilon)=(\mathcal{V})$, so that $\mathcal{P}(\mathcal{B}(\to_\epsilon)) = \emptyset$ and $\mathrm{BFDR}(\mathcal{B}(\to_\epsilon))=0$.
On the other hand, if some $q_{ij}=1$, choose $\epsilon<1$ sufficiently close to one so that $q_{ij}>\epsilon$ holds only for those pairs with $q_{ij}=1$.
For any between-block directional claim $(i,j) \in\mathcal{P}(\mathcal{B}(\to_\epsilon))$, we have $q_{ij}=1$ and hence $q_{ji}=0$.
Therefore, $\mathrm{BFDR}(\mathcal{B}(\to_\epsilon))=0$.
Thus, $\mathfrak{B}_\alpha\neq\emptyset$ for every $\alpha\in(0,1/2)$.

\textbf{Nestedness of attainable claim sets}:
Let $1/2\leq \epsilon'\leq \epsilon<1$.
Then
\begin{equation}
    \{(i,j) \mid q_{ij}>\epsilon\} \subseteq \{(i,j):q_{ij}>\epsilon'\}.
\end{equation}
Hence, every blockwise ranking with respect to $\to_\epsilon$ is also a blockwise ranking with respect to $\to_{\epsilon'}$.
Since $\mathcal{B}(\to_{\epsilon'})$ is the finest blockwise ranking induced by $\to_{\epsilon'}$, it refines $\mathcal{B}(\to_\epsilon)$.
Consequently,
\begin{equation}
    \mathcal{P}\bigl(\mathcal{B}(\to_\epsilon)\bigr) \subseteq \mathcal{P}\bigl(\mathcal{B}(\to_{\epsilon'})\bigr).
\end{equation}
Thus, the claim sets associated with elements of $\mathfrak{B}$ are totally ordered by inclusion.

\textbf{Existence of a maximizer}:
Since $\mathcal{V}$ is finite, the number of ordered partitions of $\mathcal{V}$ is finite; hence the set $\mathfrak{B}$ and its subset $\mathfrak{B}_\alpha$ are finite.
As $\mathfrak{B}_\alpha$ is nonempty, a maximizer of $|\mathcal{P}(\mathcal{B})|$ over $\mathfrak{B}_\alpha$ exists.

\textbf{Uniqueness of the maximizer}:
Suppose that $\mathcal{B}$ and $\mathcal{B}'$ both maximize $|\mathcal{P}(\cdot)|$ over $\mathfrak{B}_\alpha$.
By the nestedness just shown, either $\mathcal{P}(\mathcal{B})\subseteq\mathcal{P}(\mathcal{B}')$ or $\mathcal{P}(\mathcal{B}')\subseteq\mathcal{P}(\mathcal{B})$.
Since both claim sets have the same size, they must be equal: $\mathcal{P}(\mathcal{B})=\mathcal{P}(\mathcal{B}')$.
Therefore, $\mathcal{B}=\mathcal{B}'$, and the maximizer $\mathcal{B}_\alpha$ is well-defined.

\textbf{Monotonicity in the BFDR level}:
Finally, let $0<\alpha\leq\alpha'<1/2$.
Since $\mathfrak{B}_\alpha \subseteq \mathfrak{B}_{\alpha'}$, and $\mathcal{B}_{\alpha'}$ maximizes $|\mathcal{P}(\mathcal{B})|$ over the larger admissible set, we have 
\begin{equation}
    |\mathcal{P}(\mathcal{B}_\alpha)| \leq |\mathcal{P}(\mathcal{B}_{\alpha'})|.
\end{equation}
Together with the nestedness of attainable claim sets, this implies $\mathcal{P}(\mathcal{B}_\alpha) \subseteq \mathcal{P}(\mathcal{B}_{\alpha'})$. \qed

\putbib[refs-BIBT]
\end{bibunit}

\end{document}